\documentclass[10pt,twocolumn]{article}

\usepackage{graphicx}
\usepackage{amsfonts}
\usepackage{amsmath}
\usepackage{amssymb}
\usepackage{mathrsfs} % cursive letters
\usepackage{lipsum}
\usepackage{color}
\usepackage{siunitx}
\usepackage{authblk}
\usepackage[margin=2cm]{geometry}
\usepackage{mathtools, cuted}

\usepackage{bm}
\usepackage{amsmath}

\title{Withdrawing a solid from a bath: how much liquid is coated?}
\author[1]{Emmanuelle Rio}
\author[1]{Fran\c{c}ois Boulogne}
\affil[1]{Laboratoire de Physique des Solides, CNRS, Univ. Paris-Sud, Universit\'e Paris-Saclay, Orsay 91405, France}

\date{\today}
\begin{document}

\twocolumn[
    \begin{@twocolumnfalse}
        \maketitle
           \begin{abstract}
               A solid withdrawn from a liquid bath entrains a film.
               In this review, after recalling the predictions and results for pure Newtonian liquids coated on simple solids, we analyze the deviations to this ideal case exploring successively three potential sources of complexity: the liquid-air interface, the bulk rheological properties of the liquid and the mechanical or chemical properties of the solid.
               For these different complexities, we show that significant effects on the film thickness are observed experimentally and we summarize the theoretical analysis presented in the literature, which attempt to rationalize these measurements.
           \end{abstract}
        \tableofcontents
    \end{@twocolumnfalse}]

%%%%%%%%%%%%%%%%%%%%%%%%%%%%%%
%
% INTRODUCTION
%
%%%%%%%%%%%%%%%%%%%%%%%%%%%%%%
\clearpage
\newpage

\section{Introduction}
\begin{figure}[h!]
    \includegraphics[width=\linewidth]{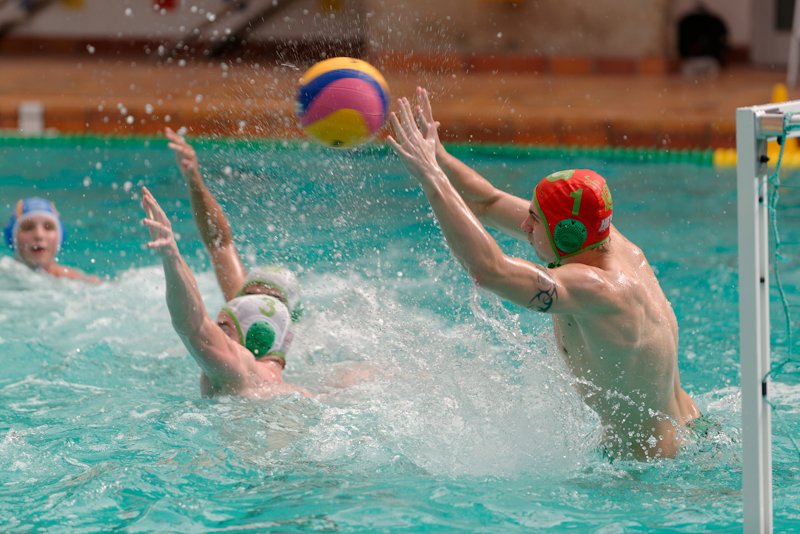}
    \caption{Jump of a water polo player to catch the ball, which entrains a water film on his body. Creative Commons By 3.0, Marie-Lan Nguyen.}
    \label{fig:intro}
\end{figure}

In our daily life, we all experienced that jumping off a swimming pool results in the formation of a water film on our body (Fig.~\ref{fig:intro}).
Beyond this curiosity, coating of solids is of crucial interest in many industrial applications.
In the middle of the twentieth century, a particular motivation came from the fabrication of photographic or motion-picture films, which are composed by almost ten layers of materials \cite{Derjaguin1964}.
The production of these films consists in coating emulsions on a flexible support hardened after cooling and drying.
The different parameters controlling the coating thickness were not understood and the development was based on empirical observations \cite{Derjaguin1964}.
More generally, liquid films on solids are important also for lubrication, rollers in printing technologies, coating of wires or optical fibers \cite{Derjaguin1964,Gutfinger1965,Wilson1982}.
Nowadays, many technological challenges rely on developing new coatings to improve material performances.
For instance, unique properties of optical glasses come from the superimposition of different layers to provide optical thin film performances \cite{Baumeister2004}.
Therefore, it is a real technological challenge to control the homogeneity and thickness of the different coated layers.

In 1922, Goucher and Ward carried out experiments to create liquid coatings on solids and to propose a dimensional analysis governing the Physics of this phenomenon \cite{Goucher1922}.
While the solution of the hydrodynamics problem for regular boundary conditions at interfaces and under certain flow regimes has been proposed in 1942 by Landau and Levich \cite{Landau1942} and then Derjaguin \cite{Derjaguin1943} in 1943, the withdrawal of a solid material from a liquid bath is still a very active subject of research, almost a century later after the first study.
The motivations to pursue studies on this subject originate from the variety and the complexity of liquid and solid properties, which are crucial to rationalize the quantity of liquid deposited on the solid.

In this review, we explore studies done during the past century on the coating of plates and fibers.
First, we briefly recall the physical mechanisms of film coating with pure Newtonian liquids on smooth solids in condition of total wetting in several hydrodynamic limits, which have been treated thoroughly by theoreticians and experimentalists.
We also present a summary of the experimental techniques that have been used.
To account for the challenges leading the actual research, we analyze the literature by considering three sources of complexity that may arise: the liquid-air interface, the bulk rheological properties of the liquid and the mechanical or chemical properties of the solid.
We show that the properties of bulk and interfaces are decisive for coating processes and that their impact on coating still contains open questions.
We choose to do not focus on the so called Bretherton problem \cite{Bretherton1961,Cantat2013} consisting of bubbles moving in capillary tubes.
Despite similarities exist between this geometry and the coating of a plate, the richness of this specific problem and the refinements that have been proposed recently would make our analysis more difficult to discern.

%%%%%%%%%%%%%%%%%%%%%%%%%%%%%%
%
% NEWTONIAN
%
%%%%%%%%%%%%%%%%%%%%%%%%%%%%%%
\section{Seminal studies: pure Newtonian liquids coated on simple solids}\label{sec:newtonian}

The coating of plates or fibers is commonly called the LLD configuration after three of the pioneering researchers who made the first theoretical derivations, namely Landau, Levich and Derjaguin.
In this section, we recall the scaling arguments in different hydrodynamics limits:
in the visco-capillary regime (the Landau-Levich regime),
in presence of gravity (the Derjaguin regime),
in presence of inertia.
We assume that the liquid is pure, Newtonian and is a perfectly wetting liquid for the solid.
For more detailed presentations of these regimes, the reader can refer to these reviews \cite{Quere1999,Ruschak1985}  and we redirect the reader to these references \cite{Quere1999,Ruckenstein2002} for full derivations as we choose to recall only the scaling laws giving the liquid film thickness.
Finally, we end this section with a presentation of experimental techniques used in the literature to study the LLD problem.

\begin{figure*}
    \begin{center}
        \includegraphics[width=16cm]{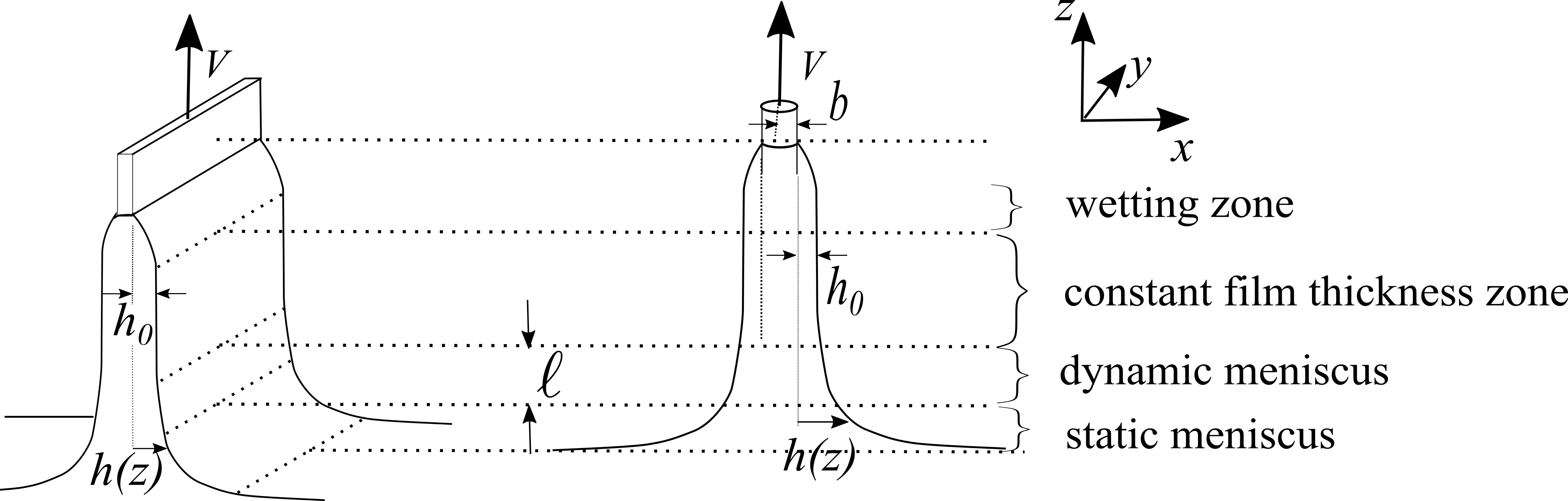}
        \caption{Coating of a solid pulled at a velocity $V$ outside a liquid bath in a plate (left) or a fiber (right) geometry. At the bottom, in the static meniscus, the interface is the same than in a static situation. The liquid is sheared in the dynamic meniscus, whose length is $\ell$. A film of constant and uniform thickness $h_0$ is formed between this dynamic meniscus and a wetting zone.}
        \label{fig:FiberVsPlate}
    \end{center}
\end{figure*}

\subsection{Shape of a static meniscus on plates and fibers}

When a solid plate is partially immersed in a liquid bath, a static meniscus rises up the solid.
In the situation of total wetting, the height $h_m$ of the static meniscus is given by the equilibrium between hydrostatic and capillary pressures.
Therefore, the meniscus height is $h_m = \sqrt{2}\ell_c$ where the characteristic length $\ell_c$ is the capillary length defined as
\begin{equation}\label{eq:cap_length}
    \ell_c = \sqrt{\frac{\gamma}{\rho g}},
\end{equation}
where $\gamma$ is the liquid-air surface tension, $\rho$ the liquid density and $g$ the gravity acceleration.
If the plate is replaced by a thin fiber, the capillary pressure has two contributions, respectively due to the vertical curvature, which is set by the capillary length $\ell_c$, and  to the azimutal curvature set by the radius $b$ of the fiber.
To quantify the relative importance of both curvatures, the Goucher number \cite{Goucher1922} is defined as the ratio of these two length scales as 
\begin{equation}\label{eq:goucher}
    {\rm Go} = \frac{b}{\ell_c}.
\end{equation}
For small Goucher numbers, the curvature due to the capillary length is negligible and the height of the meniscus becomes $\sqrt{2}b$.
In the following, we will refer to fibers when the radius of curvature of the solid material is such as $b \ll h$, \textit{i.e.} $\rm{Go}\ll 1$.

\subsection{The Landau-Levich regime}
Let us consider that a plate or a fiber is pulled out of the liquid of viscosity $\eta$ at a constant velocity $V$ and present the seminal experimental and theoretical results.
From a dimensional analysis, Goucher and Ward showed that the relevant numbers to describe the film thickness $h_0$ on fibers are $h_0/b$ and the capillary number
\begin{equation}\label{eq:Ca}
    {\rm Ca} = \frac{\eta V}{\gamma},
\end{equation}
which compares the viscous and the capillary forces \cite{Goucher1922}.
From experimental measurements on a flat surface, Morey obtained that the film thickness scales as $V^{0.63}$ \cite{Morey1940}.

A theoretical derivation of the film thickness in the stationary regime has been proposed by Landau and Levich in 1942 \cite{Landau1942}.
This description relies on the decomposition of the film in four domains depicted in Fig.~\ref{fig:FiberVsPlate}.
Close to the bath, the liquid interface position nearly corresponds to the \emph{static meniscus}.
Above the bath and along the solid, a film of \emph{constant thickness} $h_0$ is coated on the solid ending at the \emph{wetting zone}, which has a lengthscale comparable to the capillary length.
The domain between the static meniscus and the flat film is a transition zone called the \emph{dynamic meniscus} in which the liquid is sheared.
The thickness profile of the film in this domain is denoted $h(z)$ as shown in Fig.~\ref{fig:FiberVsPlate}.
The wetting zone does not enter in the original description proposed by Landau and Levich and we discuss its importance in Sec.~\ref{sec:solid}.

From physical analysis, the flow in the dynamic meniscus controls the entire dynamics of the liquid film and, in particular, its thickness.
At sufficiently small pulling velocity, inertia and gravity can be neglected.
The shear stress $\tau$ is essentially viscous and scales as $\tau = \eta \dot\gamma \sim \eta V / h_0$.
The resulting viscous stress gradient is balanced by the capillary pressure gradient $\frac{\partial}{\partial z} \left( \gamma \frac{\partial^2 h}{\partial z^2}\right)$
\begin{equation}\label{eq:ell_stress_balance}
    \eta  V / h_0^2 \sim \gamma h_0 / \ell^3,
\end{equation}
where $\ell$ is the length of the dynamic meniscus obtained by matching the curvatures of the static and the dynamic menisci.
The curvature of the dynamic meniscus scales as $ h_0 / \ell^2$. 
For a plate, the curvature of the static meniscus is $ 1 / \ell_c$ such that
\begin{equation}\label{eq:ell}
    {\ell^p} \sim \sqrt{h_0 \ell_c}.
\end{equation}
For a fiber, as explained in the previous paragraph, the typical curvature of the static meniscus is $ 1 / b$, which leads to
\begin{equation}\label{eq:ell_fiber}
    {\ell^f} \sim \sqrt{h_0 b}.
\end{equation}
We introduce the dimensionless thickness $\tilde{h}_0=h_{0}/\ell_c$ in the plate geometry and $\tilde{h}_0=h_0/b$ in the fiber geometry.
The combination of Eq.~\eqref{eq:ell} with the balance between capillarity and viscous dissipation given by Eq.~\eqref{eq:ell_stress_balance} leads to the scaling of the film thickness
\begin{equation}\label{eq:ScalingLL}
    \tilde{h}_0 \sim \rm{Ca}^{2/3}.
\end{equation}

The missing prefactor in Eq.~\eqref{eq:ScalingLL} can be calculated using an asymptotic matching of the curvature in the static meniscus \cite{Landau1942} with a numerical resolution that provides the prefator.
This full calculation gives
\begin{subequations}
    \begin{align}
        \tilde h_{\rm LLD}^p&= 0.94 \,{\rm Ca}^{2/3} \qquad \mbox{for plates,} \label{eq:LLD_plates} \\
        \tilde h_{\rm LLD}^f &= 1.34 \,{\rm Ca}^{2/3} \qquad \mbox{for fibers}, \label{eq:LLD_fibers}
    \end{align}
\end{subequations}
where we denote $h_{\rm LLD}$ the film thickness predicted in the Landau-Levich regime normalized by the capillary length or the fiber radius for plates and fibers respectively.

To present some order of magnitudes, let us take the example of a silicon oil of density $\rho=800$ kg/m$^3$, surface tension $\gamma=20$ mN/m, and viscosity $\eta=10^{-2}$ Pa$\cdot$s.
For a velocity of $V=1$ mm/s ($\textrm{Ca}=3.3 \times 10^{-5}$), the expected thickness is $h_0=6$ $\mu$m for a plate and $h_0=140$ nm for a fiber whereas the length of the dynamic meniscus is expected to be respectively $\ell^p \approx 190$ $\mu$m and $\ell^f \approx 3$ $\mu$m.
The length of the dynamic meniscus is therefore much smaller than the one of the static meniscus.

\begin{figure}
    \includegraphics[width=\linewidth]{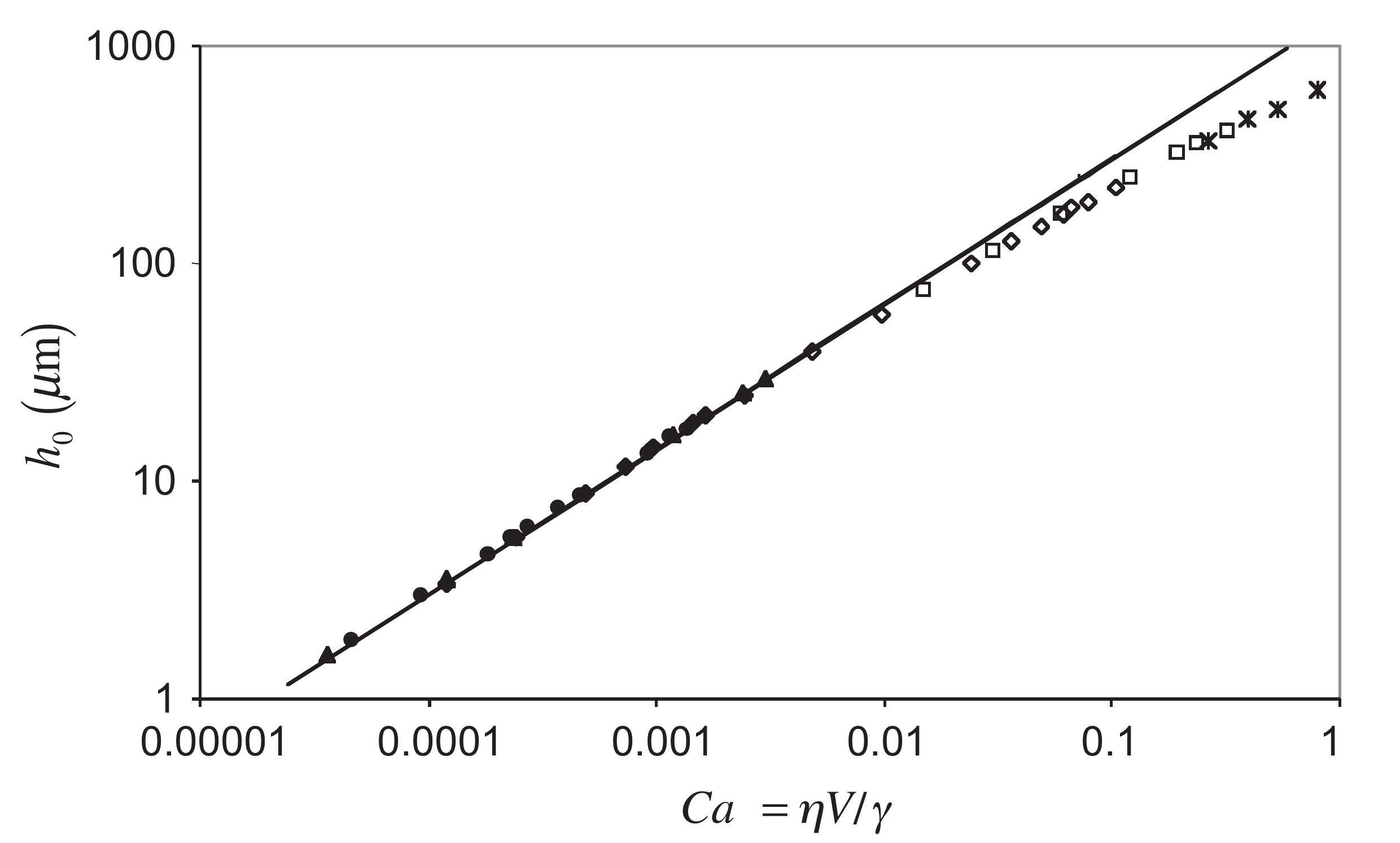}
    \caption{Experimental verification of the Landau Levich prediction (Eq.~\eqref{eq:LLD_plates}) on a plate for a pure liquid (here, silicon oils of surface tension $\gamma = 20$ mN/m).
    The thickness is plotted as a function of the capillary number and the solid line represents equation \eqref{eq:LLD_plates}.
    Figure extracted from reference \cite{Maleki2011}.}
    \label{fig:LLDPureLiquid}
\end{figure}

Many experiments have been carried out to ensure the validity of Eqs.~\eqref{eq:LLD_plates} and \eqref{eq:LLD_fibers}.
However, as we explain in the following sections, small discrepancies can appear due to more subtle bulk or interfacial effects.
Thus, only few published works are accurate enough to validate this result \cite{Quere1999, Snoeijer2008,Maleki2011,Ouriemi2013,Delacotte2012,Seiwert2011}.
An example is shown in Fig.~\ref{fig:LLDPureLiquid} \cite{Maleki2011}.

\subsection{The Derjaguin regime}\label{sec:derjaguin_regime}
Gravity has been neglected in the previous derivations and we expect that gravity is negligible only if $\rho g \ll \gamma \frac{h}{\ell^3}$.
This condition together with Eq.~\eqref{eq:ell} leads to a limit $\textrm{Ca}^{1/3} \sim 1$  above which, gravity is not negligible.
Considering a small effect gravity for ${\rm Ca}^{1/3} < 10^{-1}$, we expect that the Landau-Levich prediction is valid for ${\rm Ca} < 10^{-3}$.
Thus, gravitational effects appear at large pulling velocities and this regime is called the Derjaguin regime \cite{Derjaguin1943} and an experimental evidence is observed in Fig.~\ref{fig:LLDPureLiquid} where the deviation from the LLD law is visible above ${\rm Ca}\sim 10^{-2}$.

If gravity dominates capillarity, the balance between viscosity and gravity gives $\eta V / h_0 \sim \rho g h_0$.
Introducing the definitions of the capillary number \eqref{eq:Ca} and the capillary length \eqref{eq:cap_length}, we obtain for a plate
\begin{equation}\label{eq:GravityRegime}
    h_0 \sim \ell_c \,{\rm Ca}^{1/2}.
\end{equation}
The transition between the Landau-Levich regime (Eq.~\eqref{eq:LLD_plates}) and the Derjaguin regime  has been studied in detailed by White and Tallmadge \cite{White1967}.
Their incorrect approximation \cite{Krechetnikov2005} of the normal stress was corrected by Spiers \textit{et al} \cite{Spiers1974}.
De Ryck also proposed a scaling in the visco-gravitational regime \cite{DeRyck1998a}.
These different theories give slightly different predictions.
Most of them compare reasonably well with experimental data \cite{White1967}.
Note that in presence of gravity, a maximum can be exhibited when the thickness $\tilde{h_0}$ is plotted versus Ca. This striking feature has been observed both numerically \cite{White1966,White1967} and experimentally \cite{Spiers1974} but no qualitative explanation is given by the authors. 
In 1982, Wilson \cite{Wilson1982} recovered the LLD regime by a proper expansion of the different terms in the Navier-Stokes equation instead of neglecting a well-chosen term in the different regimes.

\subsection{Effect of inertia}\label{sec:NewtonianInertia}
In the Landau-Levich problem, inertia appears in the dimensionless form of the Navier-Stokes equation with the Weber number ${\rm We}=\frac{\rho V^2 \ell_c}{\gamma}= {\rm Re}\, {\rm Ca}$ 
\cite{Campana2010, Krechetnikov2006,Krechetnikov2010}. 
This appears naturally if the different parameters are made dimensionless using the following scales: the pulling velocity 
$V$ for velocities, the capillary length $\ell_c$ and $h_0$ respectively for the the lengths in the $z$ and $x$ directions 
and $\gamma/\ell_c$ for the pressure.
The Weber number is indeed the right parameter to quantify inertia since it compares inertia to capillarity.
The inertia contributes to the film thickening because the plate velocity is in the upward direction and has to be compared
to the capillary suction, which is the main thinning mechanism.
In the literature, the Weber number is not always the parameter exhibited by the authors because they use different definitions of the Reynolds number \cite{DeRyck1998a,Koulago1995,Soroka1971}.
In particular, in some cases, the Reynolds number is defined using the film thickness, which depends on the pulling velocity.
This leads to a slightly different prefactor of the inertial term in the Navier-Stokes equation.

The inertial regime at small capillary numbers has been well described by de Ryck and Qu\'er\'e \cite{DeRyck1998a}. The regime at large $\textrm{Ca}$ has been explored by different authors.
Different numerical resolutions are proposed by Soroka and Tallmadge \cite{Soroka1971} or White and Tallmadge \cite{White1966} leading to comparable thinning of the coated film compared to the Landau-Levich solution.
Kizito \textit{et al.} measure a saturation of the thickness at high capillary numbers \cite{Kizito1999}.
Esmail and Hummer \cite{Esmail1975} propose that these different models only describe partially the experimental data and that the $y$-direction (Fig.~\ref{fig:FiberVsPlate}) must be taken into account to describe the entire inertial problem.
The comparison of their model is in good agreement with the different data set and their model also describes very well the data obtained by Spiers \textit{et al} \cite{Spiers1974}.

\subsection{Experimental techniques}

\begin{table}
    \centering
    \begin{tabular}{|p{3cm}|p{1.9cm}|p{1.9cm}|}
        \hline
        Techniques & Plates & Fibers \\
        \hline
        Weighing &
        \cite{Morey1940,Krechetnikov2005,Ouriemi2013} \cite{Maillard2014,Maillard2015,Maillard2016} & \cite{White1966,DeRyck1994,DeRyck1998,Shen2002,Quere1997,OuRamdane1997} \\
        \hline
        Volume variation &  &\cite{Carroll1973,Quere1989}  \\
        \hline
        Galvanometer & \cite{VanRossum1958,Gutfinger1965,Spiers1974}& \\
        \hline
        Micrometric screw & \cite{VanRossum1958,Kizito1999} & \\
        \hline
        Spectroscopy Interferometry Reflectometry  & \cite{Darhuber2000,Qu2002,Snoeijer2008} \cite{Delacotte2012,Scheid2010,Seiwert2011}& \\
        \hline
        Optical distortion & \cite{Snoeijer2006,Snoeijer2008}  & \\
        \hline
        Imaging & \cite{VanRossum1958,Qu2002} \cite{Kajiya2013,Kajiya2014}& \\
        \hline
        Particle Image Velocimetry (PIV) & \cite{Kizito1999,Mayer2012} \cite{Maillard2014,Maillard2015,Maillard2016}& \\
        \hline
    \end{tabular}
    \caption{References to experimental techniques used in the literature for the two geometries to measure the film thickness and  the velocity field.}\label{tab:techniques}
\end{table}

In this Section, we present a synthetic view of the different experimental techniques of the studies cited in this review.
These techniques are summarized in Tab.~\ref{tab:techniques}, which is not intended to be comprehensive.

In several studies, the film thickness is measured by weighing the withdrawn solid.
This simple technique presents the advantage to work for any kind of non-volatile liquid and has been used in early studies \cite{Morey1940,White1966,DeRyck1994,Quere1997,OuRamdane1997,DeRyck1998} as well as recent ones \cite{Shen2002,Krechetnikov2005,Ouriemi2013,Maillard2014,Maillard2015,Maillard2016}.
However, an assumption is made on the coating uniformity.
This assumption can be verified by withdrawing solids of different sizes to check that boundary effects are negligible.
A variation of the weighing method consists in measuring the volume variation of the liquid bath.
This approach has been mainly used for fibers for which a long length of solid can be pulled.

The measurement of the local film thickness can be achieved by different methods.
Early studies used a galvanometer, from which the film thickness is deduced, knowing the liquid conductivity \cite{VanRossum1958,Gutfinger1965,Spiers1974}.
For non-conductive liquid, the alternative consists in using a micrometric screw pushing a needle toward the liquid-air interface \cite{VanRossum1958,Kizito1999}.
The contact of the needle with this interface is then  detected by eye.
As they require some adjustments, both techniques are often set up on experimental apparatus where a rotating belt mimics an endless solid.

More recently, the film thickness has been measured by the mean of optical interferometry.
The first technique is monochromatic interferometry with a laser beam \cite{Qu2002}.
To obtain an absolute thickness measurement, fringes must be counted until a reference is obtained  by a displacement toward the contact line.
Then, with an estimate of this final thickness, typically a quarter of the wavelength, the film thickness can be estimated retrospectively.
To circumvent fringes counting, more recent studies used white light spectrometry \cite{Snoeijer2008,Scheid2010,Seiwert2011,Delacotte2012}.
A white light is shed on the film and the combination of interferences for different wavelengths allows to retrieve the absolute film thickness instantaneously.
A second optical method has been developed by Snoeijer \textit{et al.} \cite{Snoeijer2006,Snoeijer2008}.
A wire is placed parallel to the coated film between the film and the camera, so that the wire and its reflection at the surface of the film are imaged together.
Then, by the mean of optical geometry, the profile of the film thickness is calculated from the deformation of the reflection of the wire.

In addition, direct visualization provides also interesting information \cite{Mayer2012}.
In 1958, Van Rossum set up a shadowgraphy technique to image the film thickness \cite{VanRossum1958}.
Images have been also used to render the shape of menisci \cite{Qu2002} and the motion of contact lines \cite{Kajiya2014}.
The flow visualization has been attempted qualitatively either by addition of particles or small bubbles \cite{VanRossum1958,Groenveld1970} and also by injection of a dye \cite{Lee1972a}.
The quantitative velocity field in the liquid can be inspected by Particle Image velocimetry (PIV) \cite{Maillard2014,Maillard2015}.
The liquid is seeded with micron-size particles illuminated by a laser sheet \cite{Mayer2012} and images are recorded with a high speed camera.
Then, PIV softwares are used to calculate the displacement field from image correlation.

%%%%%%%%%%%%%%%%%%%%%%%%%%%%%%
%
% SURFACTANT
%
%%%%%%%%%%%%%%%%%%%%%%%%%%%%%%

\section{Effect of the liquid-air interfacial properties}\label{sec:surfactant}

Experimentally, two deviations from the Landau-Levich predictions are observed in presence of surfactants.
As detailed in the next paragraph, the first one concerns the pre-factor only and the second one the scaling $\tilde{h}_0 \sim \textrm{Ca}^{2/3}$ itself, where $\textrm{Ca}$ is calculated using the equilibrium surface tension.

\begin{table*}

    \begin{tabular}{|p{3cm}|p{5cm}|p{5cm}|p{3cm}|}
        \hline
        Reference & Surfactant & Concentrations & Geometry \\
        \hline
        Shen \textit{et al} \cite{Shen2002} & BSA (Bovine Serum Albumine) &  10$^{-4}$ \%-0.16 \% & Fiber \\
        & Triton X100 & 0.06-625 cmc & Fiber \\
        & SDS (Sodium Dodecyl Sulfate) &  0.04-83 cmc & Fiber \\
        Qu\'er\'e \textit{et al.} \cite{Quere1997} & SDS &  8 cmc & Fiber \\
        Qu\'er\'e \textit{et al.} \cite{Quere1999} & SDS &  0.01-10 cmc & Fiber \\
        Krechetnikov \textit{et al.} \cite{Krechetnikov2005} & SDS &  0.25-1 cmc & Plate \\
        Delacotte \textit{et al.} \cite{Delacotte2012} & C$_{12}$E$_{6}$ (Hexaethylene Glycol Monododecyl Ether) &  0.5-50 cmc & Plate \\
        & DTAB (dodecyl trimethyl ammonium bromide) &  0.5-25 cmc & Plate \\
        & DeTAB (decyl trimethyl ammonium bromide) &  7.5 and 15 cmc & Plate \\
        Mayer \textit{et al.} \cite{Mayer2012} & SDS &  0.25-0.5-1-5 cmc & Plate \\
        \hline
    \end{tabular}
    \caption{Surfactant molecules and concentrations used in plate or fiber geometry to measure the coated thickness.}\label{tab:systems}
\end{table*}

\begin{figure}
    \includegraphics[width=\linewidth]{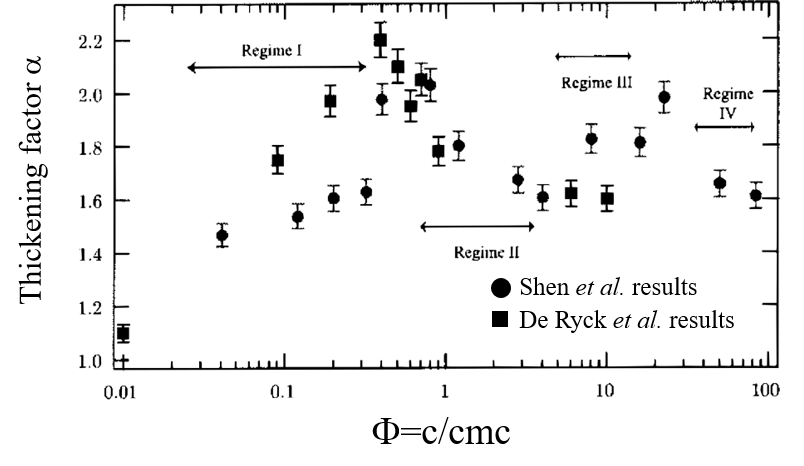}
    \caption{Experimental measurements of the thickening factor versus the surfactant concentration obtained in two different studies, namely \cite{Shen2002} and \cite{Quere1997}. Figure from \cite{Shen2002}.}
    \label{fig:Alpha}
\end{figure}

The different experimental systems available in the literature are summarized in Tab.
\ref{tab:systems}.
In some experiments, the scaling law ${\rm Ca}^{2/3}$ is preserved \cite{Shen2002,Quere1997,Krechetnikov2005,Delacotte2012,Scheid2010} and only the prefactor varies.
Thus, a thickening factor $\alpha=\frac{\tilde{h}_0}{\tilde{h}_{\rm LLD}}$ can be conveniently introduced.
An overview of the different experiments available in 2002 is shown in Fig.~\ref{fig:Alpha}.
In other sets of experiments, the scaling law $h_0 \propto \textrm{Ca}^{2/3}$ is no more observed \cite{Quere1997,Delacotte2012,Champougny2015}. Most of the time, deviations occur at capillary numbers close to ${\rm Ca} \simeq 10^{-4}$, where the thickness increases slower than predicted by the scaling.

From the hydrodynamics point of view, the replacement of a pure liquid by a solution containing surfactants, particles or polymers mainly affects the liquid-air boundary condition through the mechanical description of the interface.

\subsection{The boundary condition at the liquid-air interface}

At a liquid-air interface, the boundary condition is the stress continuity.
In the well-known case of a pure liquid, the boundary condition in the direction tangential to the interface is simply a no-stress boundary condition: $\left. \eta \frac{\partial u_z}{\partial x}\right|_{x=h(z)}=0$ where $u_z$ the vertical velocity field, $x$ the horizontal direction perpendicular to the plate and $h(z)$ is the interface profile (Fig.~\ref{fig:FiberVsPlate}).

In presence of surface active molecules, an additional stress due to the mechanical properties of the interface (surface rheology) can appear at the interface which affects the boundary condition \cite{Langevin2014}.
The writing of this additional term in the general case implies a tensorial description of the interfacial stress  \cite{Edwards1991,Fuller2012,Kistler1997}.
Consequently, in the following, we present this boundary condition only in the specific geometry of a film pulled out of a bath.
In addition, we assume a translation symmetry in the $y$ direction such that the problem can be treated in two dimensions and we also assume that the conditions to apply the lubrication approximation are met.

The following equation is an attempt to take into account the different possible stress contributions:
\begin{equation}\label{eq:BC}
    \left. \eta \frac{\partial u_z}{\partial x}\right|_{x=h(z)}=    \overbrace{K(E,G)}^{\textrm{solid elasticity}} + \overbrace{\frac{\partial \gamma}{\partial z}}^{\textrm{Marangoni}}+\overbrace{\mu^{\star} \frac{\partial^2 u_s}{\partial z^2}}^{\textrm{surface viscosity}},
\end{equation}
where $\mu^{\star}$ is a surface viscosity, which is a combination of the shear and dilational surface viscosities $\mu_s$ and $\kappa_s$ \cite{Scheid2010} ($\mu^{\star}=\mu_s+\kappa_s$), and  $E$ and $G$ are respectively the dilational and shear elasticities.

The first term on the right-hand side of  Eq.~\eqref{eq:BC} describes the solid surface elasticity.
This term depends on two parameters, the shear and the dilational surface elasticities and appears for complex objects at liquid-air interfaces such as polymers or particles.

The second term on the right-hand side of Eq.~\eqref{eq:BC} is the Marangoni stress, due to a gradient of surface tension.
For solutions of surface active molecules, any variation of the surface tension due to advection, surface diffusion or exchanges between surface and bulk can lead to a surface tension gradient.
As we will see in the following, this is the contribution of the \emph{Gibbs-Marangoni elasticity} that measures the ability of a surface to sustain surface tension gradients.
In pure liquid, surface tension gradients may originate from temperature gradients.
For instance, evaporation can induce such temperature gradients \cite{Qu2002}.
Nevertheless, in this section, we choose to focus on the effect of surface active molecules.

The third term on the right-hand side of Eq.~\eqref{eq:BC} is due to surface viscosity. This additional stress is due to the dissipation at the interface, either because of surface/volume exchanges or because of molecular friction \cite{Cantat2012}.
The surface viscosity thus makes the system more complex through the apparition of a second derivative of $u_s$ in the boundary condition.

\subsection{Film thickness for different interfacial boundary condition}

In the different studies available in the literature to model these deviations, some terms on the right-hand side of Eq.~\eqref{eq:BC} are neglected depending on the approximations or hypothesis made by the authors. 
In the following,  we focus on how and when the consideration of each term in the right-hand side of Eq.~\eqref{eq:BC} can explain the deviations to the Landau-Levich law.

\subsubsection{The so-called rigid limit} \label{sec:rigid}

\begin{figure}
    \centering
       \includegraphics[width=\linewidth]{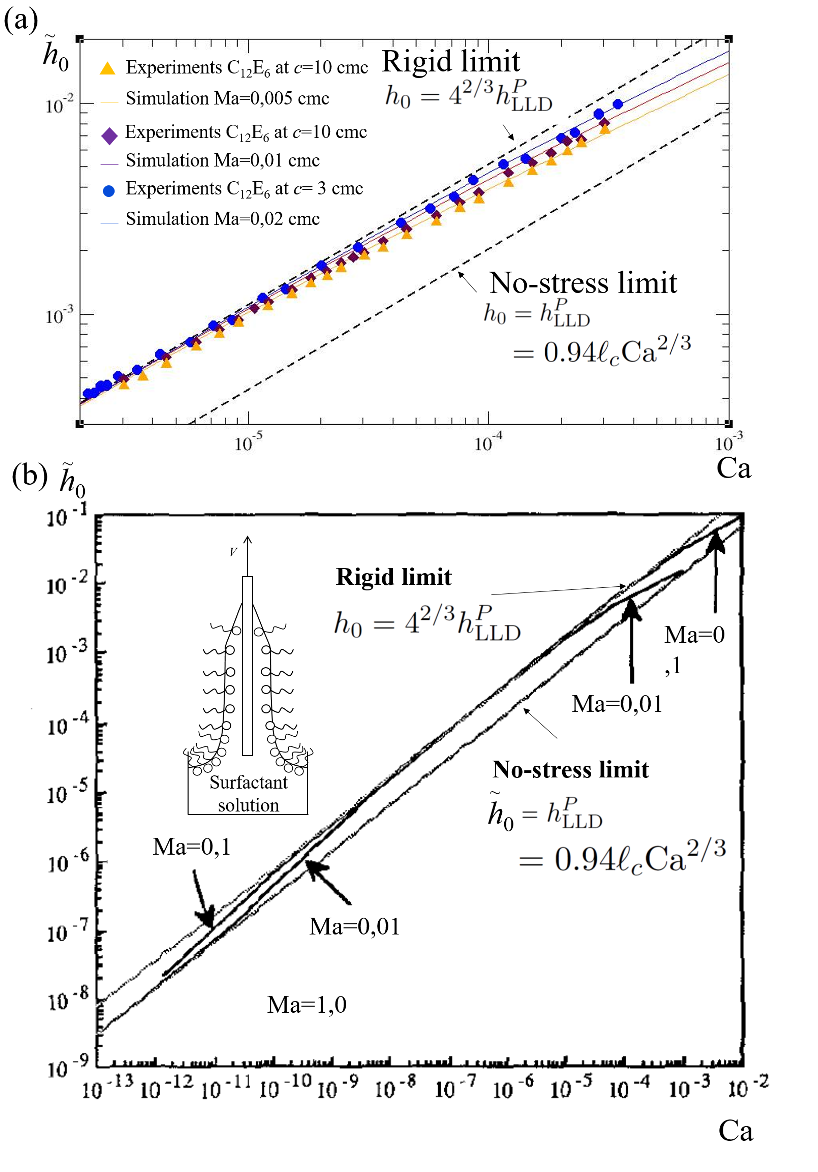}
       \caption{Both graphics represent the dimensionless thickness $\tilde{h}_0$ versus the capillary number ${\rm Ca}$ in log-log scale.
The straight lines (solid in (a) and dashed in (b)) represent respectively the Landau-Levich prediction given by Eq.~\eqref{eq:LLD_plates} (lower straight lines) and the rigid limit (upper straight lines), which is $4^{2/3}$ times larger.
(a) Experimental data obtained by pulling a plate out of a solution of C$_{12}$E$_6$ at different concentrations $c$ measured in unity of the cmc.
The data are fitted by numerical simulations of the thickness of the liquid film using the Marangoni number as a fitting parameter.
(b) Numerical simulations of the thickness of the liquid film entrained by a plate pulled out of a surfactant are plotted for different values of the Marangoni number. 
(a) is adapted from \cite{Champougny2015} and (b) from  \cite{Park1991}.
         }
        \label{fig:Marangoni}
\end{figure}

For situations in which one of the right-hand term of Eq. \eqref{eq:BC} is large compared to the viscous dissipation on the left-hand side, the interfacial stress can dominate such that the boundary condition at the liquid-air becomes $V(h)=0$.
This limit is often referred as the rigid limit because the boundary condition is the same as the one at a liquid/solid interface with the same velocity.
In this limit, the Landau-Levich model still holds (including the prefactor) provided that the velocity $V$ is replaced by $4V$ \cite{Quere1997}, leading to a maximum thickening factor $\alpha_{\textrm{max}}=4^{2/3}\approx 2.5$, i.e.
\begin{equation}\label{eq:LLDRigid}
    \tilde{h}_0=4^{2/3} \tilde{h}_{\rm LLD}.
\end{equation}
In Fig.~\ref{fig:Marangoni}, the straight lines have a power $2/3$ and the prefactors correspond respectively to the Landau-Levich prediction given by Eq.~\eqref{eq:LLD_plates} at the bottom and to the maximum thickness given by Eq.~\eqref{eq:LLDRigid}, which is $4^{2/3}$ larger.

\subsubsection{Effect of the solid surface elasticity}
The first term of Eq.~\eqref{eq:BC} in the LLD problem is always neglected in presence of surfactants.
An attempt to incorporate this term has been done by Homsy \textit{et al.} \cite{Ouriemi2013,Dixit2013} to describe complex interfaces, for instance in presence of particles or polymers.
These authors introduce an expression for $K(E,G)$ that depends on the bending modulus $K_B$ and they predict a scaling law for the film thickness 
\begin{equation}
	h_0 \sim \ell_e {\rm El}^{4/7},
\end{equation}
where $\ell_e=\sqrt[]{\frac{K_B}{\rho g}}$ and ${\rm El}=\frac{\eta V \ell_e^2}{K_B}$ is the elasticity number \cite{Dixit2013}.
Their prediction successfully describes the power law of their experiments with micron size fingerprint hydrophobic powder but the prefactor still requires some refinements \cite{Ouriemi2013}.

\subsubsection{Effect of the Gibbs-Marangoni elasticity}

In this section, we will focus on the effect of the second term in Eq.~\eqref{eq:BC}, \textit{i.e.} the Gibbs-Marangoni elasticity.
Qualitatively, the presence of surfactants leads to a Marangoni stress at the liquid air/interface because the expansion of the interface caused by the movement of the solid is concentrated in the dynamic meniscus \cite{Carroll1973}.
The surface concentration in surfactants is  thus larger at the bottom than at the top (inset in Fig.~\ref{fig:Marangoni}(a)).

Different authors (\cite{Park1991,Tiwari2006,Krechetnikov2006, Campana2010,Campana2011,Champougny2015} among others) proposed numerical simulations of the effect of Gibbs-Marangoni elasticity by integrating the Stokes equation together with the boundary condition given by Eq. \eqref{eq:BC} containing only the second term on the right-hand. Different simplifications are reviewed in the following.

\paragraph{Model of insoluble surfactants in absence of surface diffusion}

In presence of insoluble surface active objects, the surface concentration $\Gamma(z,t)$ affects the surface tension through the combination of advection and diffusion as stated by the mass conservation for the surface active molecules
\begin{equation}\label{eq:SurfactantConservationInsoluble}
    \frac{\partial \Gamma}{\partial t}+\frac{\partial \Gamma}{\partial z}.u_s=D_s\frac{\partial^2 \Gamma}{\partial z^2},
\end{equation}
where $D_s$ the surface diffusion coefficient of the surface active object.

To close the problem, a state equation is needed to relate the surface tension to the surface concentration.
The Gibbs-Marangoni surface elasticity defined as $E_{\rm GM}=-\frac{1}{\Gamma} \frac{\partial \gamma}{\partial \Gamma}$ quantifies the variation of the surface tension with the surface concentration.
By integrating this definition of $E_{\rm GM}$, we obtain the following state equation:
\begin{equation}\label{eq:StateEquationInsoluble}
    \gamma(\Gamma)=\gamma_0-E_{\rm GM} \ln{\left(\frac{\Gamma}{\Gamma_0}\right)},
\end{equation}
where $\gamma_0$ is the surface tension at a surface concentration $\Gamma_0$.

In absence of surface diffusion, \textit{i.e.} $D_s = 0$ in Eq.~\eqref{eq:SurfactantConservationInsoluble}, a unique control parameter $\Lambda=\frac{{\rm Ma}}{{\rm Ca}^{2/3}}$ can be exhibited, which compares the surface and the bulk stress \cite{Tiwari2006,Champougny2015}.
In this expression, the Marangoni number is defined as ${\rm Ma}=\frac{E_{\rm GM}}{\gamma}$.
Ruckenstein \cite{Ruckenstein2002} also found this control parameter, without its explicit expression, by adding a Marangoni velocity to the pulling velocity.
In presence of large surface tension gradients, the Marangoni number is large and the dissipation mainly occurs at the interface such that the rigid limit presented in Sec.~\ref{sec:rigid} is recovered and the scaling law $\textrm{Ca}^{2/3}$ holds.
At high velocity and/or for small Marangoni numbers, no scaling law can be obtained \cite{Champougny2015}.
The thickness progressively leaves the rigid limit and get closer to the Landau-Levich limit (see experimental data in Fig.~\ref{fig:Marangoni}(a)).

To compare the simulations to experimental data, the Marangoni number can be used as a fitting parameter (Fig.~\ref{fig:Marangoni}(b)).
However, for a correct understanding, the obtained Marangoni number must be compared to a measurable parameter and such comparisons are scarce in the literature.
Champougny \textit{et al.} proposed to work in the insoluble limit (within the same limit than Park \cite{Park1991} or Campana \cite{Campana2011}) so that Eq.~\eqref{eq:StateEquationInsoluble} can be used.
%The Marangoni number thus writes $\textrm{Ma}=\frac{E_{GM}}{\gamma}$ 
and $E_{GM}$ can be compared to experimental measurements of the surface elasticity performed in a Langmuir through.
The comparison between experiments and theory is surprisingly successful since a model of insoluble surfactants is used to describe data obtained with soluble surfactants.
This is certainly because the surfactants used in this study have an adsorption time much larger than the experimental time, so that they can be considered as insoluble.

Another lead to compare numerical simulations to molecular parameters ($E_{\rm GM}$ or $\eta_s$) is to use not only the prediction of the film thickness but also the prediction of the surface velocity $u_s$, which is a direct measurement of the comparison between the stress at the liquid-air interface and in the bulk.
A systematic measurement of this surface velocity has been proposed by Mayer \textit{et al} \cite{Mayer2012}.
The authors use a direct visualization of the flow field to extract this parameter.
They obtain a surface velocity, which is comparable to the pulling velocity $V$ at small capillary number. This means that the surface rises at the maximum velocity and that the conditions to obtain the thickness predicted by Eq.~\eqref{eq:LLDRigid} are fulfilled.
At higher capillary numbers, the surface velocity decreases, in agreement with the simulations of Fig.~\ref{fig:Marangoni}(a).
They also measure a surface velocity, which decreases continuously with an increasing surfactant concentration.
A lack of direct measurement of the surface viscoelasticity prevents a quantitative comparison with their predictions.

\paragraph{Model of insoluble surfactant: effect of surface diffusion}
Park \cite{Park1991} gave a comprehensive description of the problem in the case of insoluble surfactants and in absence of surface viscosity by adding the surface diffusion.
Fig.~\ref{fig:Marangoni}(b) shows the thickness obtained from Park's simulations.
At small Ca, the thickness is close to the stress free limit (Eq.~\eqref{eq:LLD_plates}, bottom straight line in the figure).
It reaches the rigid case at intermediate Ca (Eq.~\eqref{eq:LLDRigid}, top straight line in the figure) before coming back to the stress free limit at high Ca.
The transition at high Ca has been described in details in the previous paragraph.
The transition at low Ca is due to surface diffusion that prevents surface tension gradients.
An extension of the model derived by Park including gravity has been proposed by Zhang \textit{et al.} \cite{Zhang2001}.
A reasonable agreement with experiments performed using insoluble surfactants as coating liquid is obtained.

\paragraph{Model of soluble surfactants}
A set of equations similar to \eqref{eq:SurfactantConservationInsoluble} and \eqref{eq:StateEquationInsoluble} can be introduced to describe the presence of soluble surface active agents but the exchanges between surface and bulk must be taken into account.
A flux $j$ of surface active agents between the bulk and the interface can occur, which can be added to the right-hand side of Eq.~\eqref{eq:SurfactantConservationInsoluble} leading to
\begin{equation}\label{eq:SurfactantConservationSoluble}
    \frac{\partial \Gamma}{\partial t}+\frac{\partial \Gamma}{\partial z}.u_s=D_s\frac{\partial^2 \Gamma}{\partial z^2}+ j.
\end{equation}
In addition, the surface concentration $\Gamma$ is not anymore the inverse of the area per molecule $A$, such that the Gibbs-Marangoni elasticity is now defined as $E_{\rm GM}=-\frac{1}{A} \frac{\partial \gamma}{\partial A}$.

The resolution of the hydrodynamics equations with the different effects of soluble or insoluble surface active agents represents a difficult theoretical and numerical task.
The comparison with experimental data is also complex for the following reasons.
First, the parameters describing the surface rheology, namely the surface viscosity and the surface elasticity, are difficult to measure experimentally since they depend on the surfactant concentration and on the characteristic timescale of the solicitation \cite{Langevin2014}.
Second, the source term $j$ is difficult to estimate and depends not only on diffusion and advection in the volume but also on adsorption/desorption barriers at the liquid-air interface.
Third, in the case that these parameters are fitted, the large number of degrees of freedom   possibly makes their estimations unreliable.

Numerical simulations performed by Krechetnikov and Hosmy include soluble surfactants and predict a thinning of the entrained film in presence of surfactants \cite{Krechetnikov2006}.
Nevertheless, such a thinning has never been observed experimentally, neither by the authors \cite{Krechetnikov2005} nor by others.
Similar simulations have been performed with the same hydrodynamical ingredients by Campana \cite{Campana2011}.
These simulations led to a thickening confirming the experimental results.
More recently, Krechtenikov \cite{Krechetnikov2010} actually explained that the prediction of a thinning is valid only if the bulk concentration is maintained 
constant, which may be the case at high surfactants concentration.
It seems that ending the controversy needs a complete solving of the flow field, not only in the thin film region but also in the bath \cite{Krechetnikov2010,Mayer2012}.
In these simulations, the power-law of the Landa-Levich law is preserved and only the prefactor is preserved.

\subsubsection{Effect of the surface viscosity}
\begin{figure}
    \includegraphics[width=\linewidth]{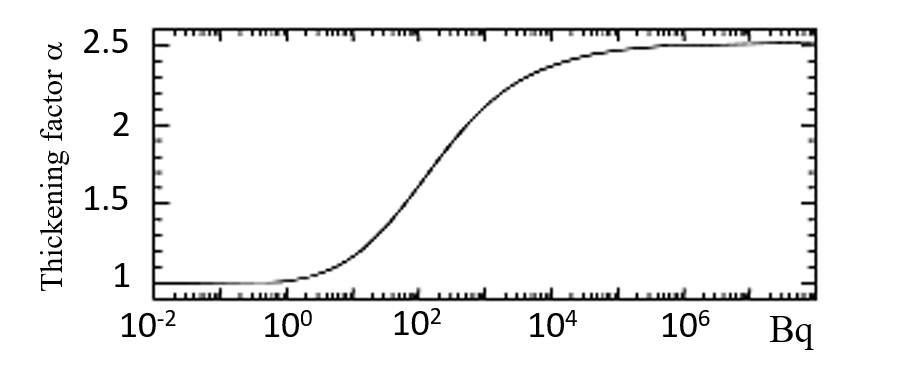}
    \caption{Prediction of the thickening factor depending on the Boussinesq number. Figure from \cite{Scheid2010}.}
    \label{fig:AlphaBq}
\end{figure}
The third term on the right-hand of Eq.~\eqref{eq:BC} corresponds to the effect of surface viscosity on the thickness of the entrained film.
The Boussinesq number compares the surface viscous stress to the bulk stress and reads $\textrm{Bq}=\frac{\eta_s}{\eta d}$ with $d$ a characteristic lengthscale.
The publications taking into account this term are scarce, but the global picture has been depicted by Scheid \textit{et al.} \cite{Scheid2010}, for which the main result is given in Fig.~\ref{fig:AlphaBq}.
By increasing the Boussinesq number, the thickening factor $\alpha$ is going from $1$ (Landau-Levich limit) to $4^{2/3}$ (rigid limit).
From an experimental point of view, it is very difficult to vary continuously the Boussinesq number bacause, most of the time, the third term on the right-hand of Eq.~\eqref{eq:BC} (surface viscosity) is most of the time negligible in front of the second term (Marangoni).
Scheid \textit{et al} identified a situation, in which the surface viscosity term is predominant by performing experiments using a solution of dTAB (decyl trimethyl ammoniumbromide).
 This surfactant has a high solubility such that experiments can be performed at very high concentrations (up to 15 times the cmc), that they compared to their numerical predictions.
At such a high concentration, the surfactant remobilization \cite{Stebe1991} is so fast that the Gibbs-Marangoni (and thus the second term on the right-hand of Eq.~\eqref{eq:BC}) becomes negligible.
No surface tension gradient occur because any depletion in surfactant at the interface is replaced at a timescale much smaller than the experimental timescale.
In this particular case, only the third term on the right-hand side of Eq.~\eqref{eq:BC} is expected to be non negligible.
The fit of the experimental data by the numerical simulations gives a surface tension $\eta_s=2 \times 10^{-5}$~Pa$\cdot$s$\cdot$m, which is consistent with typical values reported in the literature for ionic surfactants \cite{Stevenson2005,Prudhomme1995}.
With the result obtained by Campana \textit{et al.} \cite{Campana2010}, this work exhibits a second situation in which the power law $2/3$ is preserved and that predicts the thickening factor.

%%%%%%%%%%%%%%%%%%%%%%%%%%%%%%
%
% BULK
%
%%%%%%%%%%%%%%%%%%%%%%%%%%%%%%

\section{Effect of the bulk rheological properties of liquids}\label{sec:bulk}

As mentioned in Sec.~\ref{sec:newtonian} where we introduce the Landau-Levich equations, the withdrawal of a solid shears the liquid in the transition region.
For Newtonian fluids, we used the constitutive equation $\tau = \eta \dot\gamma$ where $\eta$ is a constant viscosity.
Therefore, we can expect that the bulk rheological properties play a significant role in the deposited thickness.
In this paragraph, we review results about shear-thinning, viscoelastic and yield stress fluids.
Even if the distinction between these different rheological properties is often made in the literature, the complexity of non-Newtonian fluids must be retained.
Indeed, combinations of these rheological behaviors can be encountered in experiments.

\subsection{Shear-thinning fluids}

\begin{figure}
    \includegraphics[width=\linewidth]{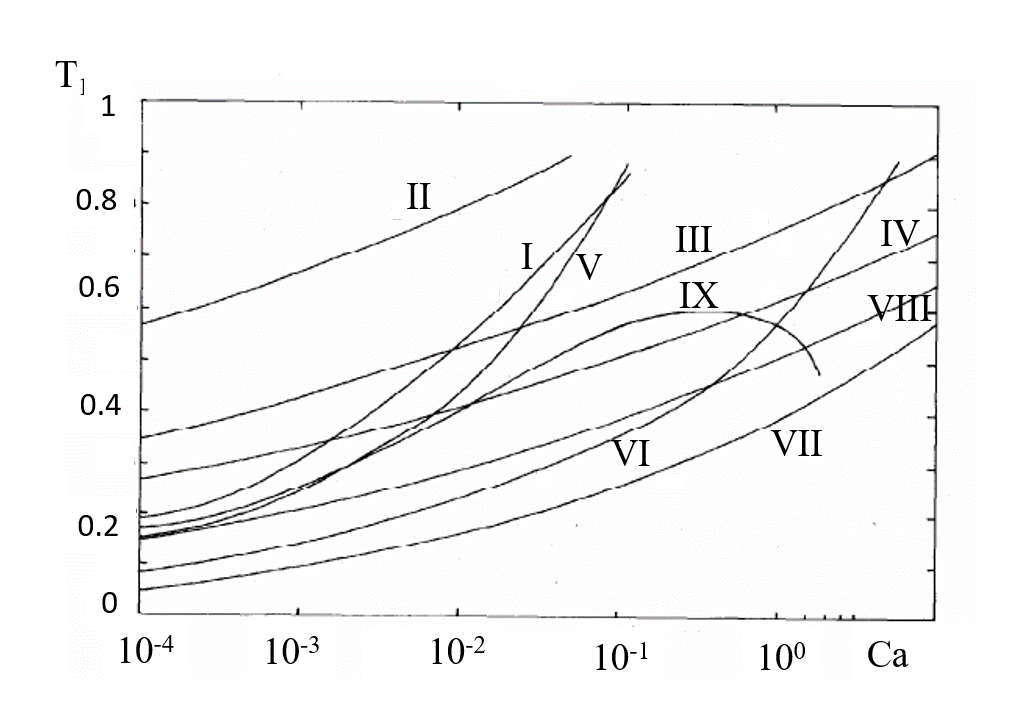}
    \caption{Comparison of different theories for $n=0.5$.
    The parameter $T_1$ is the thickness normalised by the expression given by Eq. \eqref{eq:Tekic}.
    The different plotted models are presented in the following references: I and II \cite{Gutfinger1965}, III \cite{Tallmadge1966}, IV \cite{Tallmadge1969}, V \cite{Groenveld1970}, VI and VII \cite{Tallmadge1970}, VIII \cite{Spiers1975}, IX \cite{Tekic1983}.
    The figure is extracted from \cite{Tekic1983}.
    }
    \label{fig:viscoelastic}
\end{figure}

Two constitutive equations are mainly considered to describe shear-thinning fluids \cite{Gutfinger1965,Tallmadge1970,Spiers1975,Tekic1983,Afanasiev2007,Hewson2009}.
The simplest model is the power law model that can be written as

\begin{equation}\label{eq:st_fluid}
    \tau = k \dot\gamma^n \qquad {\rm or} \qquad \eta = k \dot\gamma^{n-1},
\end{equation}
where $k$ is the consistency index and $\dot\gamma$ is the shear rate.
In this model, the low shear behavior is often non-realistic.
Indeed, a better description of the low shear behavior consists in imposing a lower viscosity limit $\eta_0$ than the power law model.
The Ellis model can be written as
\begin{equation}\label{eq:ellis}
    \eta = \frac{\eta_0}{ 1 + (\tau/\tau_{1/2})^{m-1}},
\end{equation}
where $\eta_0$ is the zero shear viscosity, $m$ a dimensionless parameter and $\tau_{1/2}$ is the shear stress value at which $\eta = \eta_0/2$.

Gutfinger and Tallmadge extended the Landau-Levich-Derjaguin theory to power law fluid \cite{Gutfinger1965}.
The capillary number is generalized for a power law constitutive equation as
\begin{equation}
    {\rm Ca}_n = \frac{k h_0^{1-n} V^n}{\gamma}.
\end{equation}
Under the assumption of small capillary number, their model suggests that the film thickness scales as
\begin{equation}
    h_0 \sim \left( \frac{k^2 V^{2n}}{\gamma^{1/2} (\rho g)^{3/2}} \right)^{1/(2n+1)}.
\end{equation}
For $n=1$, which corresponds to a Newtonian fluid of viscosity $k=\eta_s$, Eq.~\eqref{eq:ScalingLL} is recovered.
Gutfinger and Tallmadge calculated the prefactor that we omit here for the sake of simplicity and for large capillary numbers, they obtained
\begin{equation}
\label{eq:Tekic}
    h_0 \sim \left( \frac{kV^n}{\rho g} \right)^{1/(n+1)}.
\end{equation}
However, their experimental results indicates that these relations provide only the order of magnitude and they overestimate their experimental findings.

Then, Spiers \textit{et al.} noticed that the comparison of former studies by Gutfinger, Tallmadge and Groeveld shows great differences in their predictions \cite{Spiers1975}, as shown in Fig.~\ref{fig:viscoelastic}.
In their study, they proposed a new theory for power law fluids based on the constitutive equation given by Eq.~\eqref{eq:st_fluid} but also on the Ellis model.
Their experiments are performed with Carbopol in water or in a mixture of water and glycerol, and Separan in water and glycerol.
They concluded that the Ellis model provides a better estimate of the order of magnitude of the film thickness.
Qualitatively, this thickness is smaller than the thickness that would be obtained from the LLD theory for a Newtonian fluid of viscosity $\eta_0$ as the effective viscosity is smaller than this value.
Nevertheless, the model proposed by Spiers \textit{et al.} is not fully satisfying to predict the experimental results.
Such discrepancies might be inherent to the complexity of the rheological properties of these solutions as they noticed normal stresses and elastic behaviors, which are not included in the theory.
In their observations, viscoelasticity tends, in general, to reduce the film thickness.

Interestingly, Teki\'c and Popadi\'c included theoretically gravito-inertial effects in a model for power law liquids \cite{Tekic1983}.
In contrast with the previous models, they found that the coated thickness as a function of the capillary number is not monotonic and exhibits a maximum (Fig.~\ref{fig:viscoelastic}) as noticed for Newtonian fluids in presence of gravity (See Sec.~\ref{sec:derjaguin_regime}).
Further refinements have been proposed for power-law and Ellis models \cite{Afanasiev2007,Hewson2009} but also for Carreau and Olroyd-B models \cite{Afanasiev2007,Ro1995}.

\subsection{Viscoelastic liquids}
The effect of normal stresses of polymer solutions has been investigated experimentally and theoretically by De Ryck and Qu\'er\'e with fibers \cite{DeRyck1998}.
To model viscoelastic solutions of flexible polymer chains, they combined a shear thinning effect and a normal stress difference $\psi_1\dot\gamma^{2n}$.
Therefore, the constitutive equation is

\begin{equation}
    \tau = \begin{bmatrix}
        \psi_1 \dot\gamma^{2n} & \eta \dot\gamma & 0\\
        \eta \dot\gamma & - \frac{1}{2} \psi_1 \dot\gamma^{2n} & 0\\
        0 & 0 & - \frac{1}{2} \psi_1 \dot\gamma^{2n}
    \end{bmatrix}
\end{equation}
with
\begin{equation}
    \eta = \left\{
        \begin{array}{ll}
            \eta_0 & \mbox{if} \quad \dot\gamma< \dot\gamma_c, \\
             k \dot\gamma^{n-1} & \mbox{if}   \quad  \dot\gamma> \dot\gamma_c,\\
        \end{array}
    \right.
\end{equation}
where $\dot\gamma_c$ is a critical shear rate.
Separating the viscous and the normal stress, they derived scaling laws for the film thickness
\begin{subequations}
    \begin{align}
        h_0 &\sim \left( \frac{b^3 k ^2}{\gamma^2} \right)^{1/(2n+1)} V^{2n/(2n+1)}, \label{eq:shearthinning_stress_quere}\\
        h_0 &\sim \left( \frac{\psi_1 b}{\gamma} \right)^{1/(2n)} V \label{eq:normal_stress_quere},
    \end{align}
\end{subequations}
for the shear thinning effect and the normal stress effect, respectively.
Equation \eqref{eq:shearthinning_stress_quere} is obtained from the balance between the capillary pressure gradient $\gamma h_0 / \ell^{f\,3}$ and the shear stress gradient $k V^n/h_0^{n+1}$.
For Eq. \eqref{eq:normal_stress_quere}, the capillary pressure gradient is balanced with the gradient of normal stresses $\frac{\psi_1}{\ell^f} \left(\frac{V}{h_0}\right)^{2n}$.
For $n=1$ and $\psi_1=0$, the scaling for a Newtonian liquid is recovered (Eq.~\eqref{eq:ScalingLL}).

For high molecular weight polyethylen oxyde solutions in a semi-dilute regime, De Ryck and Qu\'er\'e obtained a swelling effect of a factor between 2 and 8.
From the equations combining shear-thinning and normal stress effects, they obtained a numerical solution that successfully describes their experimental observations and they attributed the swelling effect to the normal stress of the polymer solutions.
Ashmore \textit{et al.} derived a formal matched asymptotic analysis based on a similar constitutive equation to that used by de Ryck and Qu\'er\'e, that they compared to experiments performed in a roller geometry.
They obtained good agreements in the weakly elastic limit with semi-dilute solutions of polyacrylamide in a glycerol and water mixture.

Ruckenstein also proposes a scaling analysis on the coating of fibers with a polymer solution exhibiting a viscoelastic behavior \cite{Ruckenstein2002}.
The model considers a viscoelastic fluid with a characteristic relaxation timescale $\theta$.
The dimensionless number that compares this elastic timescale to the viscous dissipation timescale is the Deborah number ${\rm De} = \theta V / b $.
For large ${\rm De} / {\rm Re}$, Ruckenstein predicts
\begin{equation}
    \frac{h_0}{b} \sim \left( {\rm Ca}\, {\rm De} \right)^{1/2} = \left( \frac{\eta \theta V^2}{\gamma b} \right)^{1/2}.
\end{equation}
Thus, for elasticity dominated flow, the thickness $h_0$ scales as the velocity $V$.
This result is equivalent to the scaling obtained by de Ryck and Qu\'er\'e with Eq.~\eqref{eq:normal_stress_quere} \cite{DeRyck1998}.

\subsection{Yield stress fluids}\label{subsec:YSF}
\begin{figure*}
    \includegraphics[width=6cm]{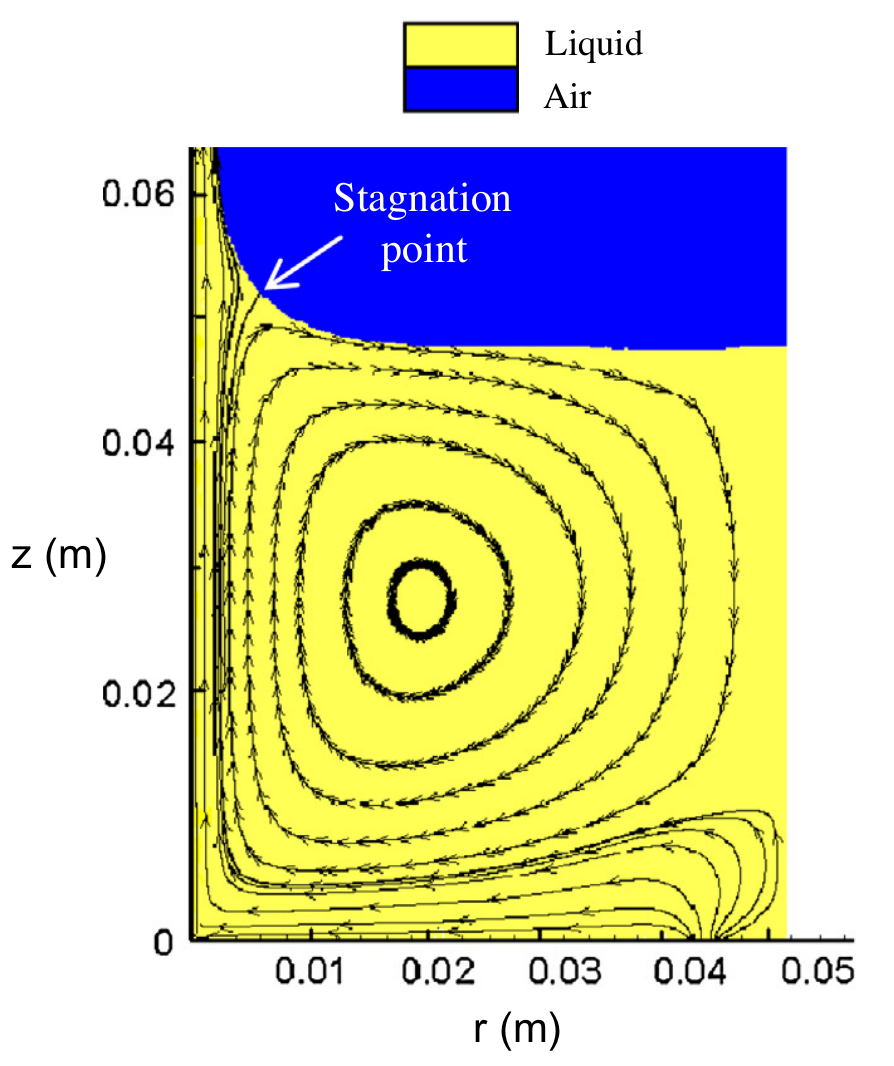}
    \includegraphics[width=6cm]{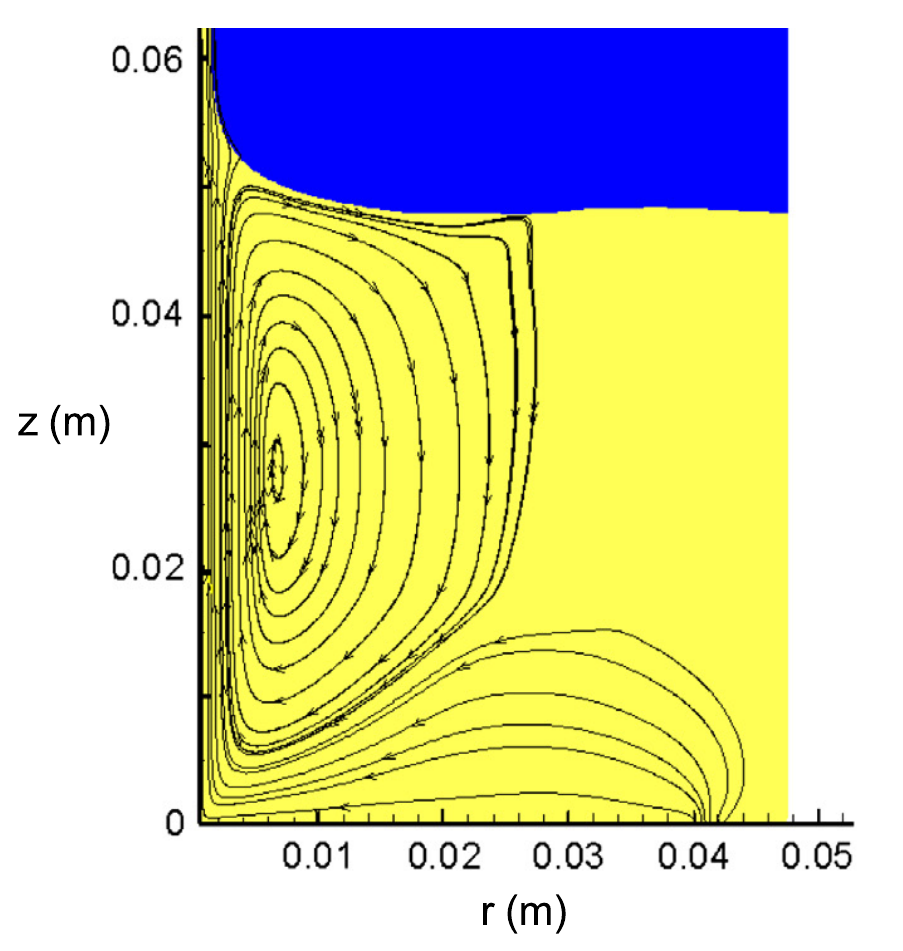}
    \includegraphics[width=6cm]{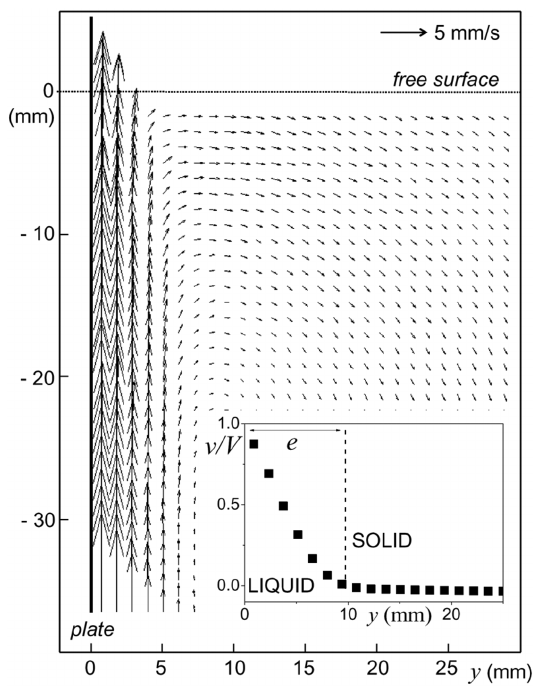}
    \caption{Liquid flow fields for plate withdrawal. (a) Numerics for a Newtonian fluid $\mu = 2.9$ Pa$\cdot$s (b) Numerics for a Bingham fluid $\mu=2.9$ Pa$\cdot$s, $\tau_0 = 4.0$ Pa.
    (a-b) are from \cite{Filali2013}.
    (c) PIV measurements for a carbopol solution with $\tau_c = 34$ Pa, $n\approx 0.35$, $k=13.9$ Pa$\cdot$s$^{-n}$ and a plate velocity $V = 15$ mm/s The figure is extracted from \cite{Maillard2014} and the rheological parameters are found in \cite{Maillard2015a}.
    }
    \label{fig:yield_stress_fluid}
\end{figure*}

Now, we consider a second class of complex fluids: the yield stress fluids.
These materials have the property to flow above a critical stress and behave as elastic solids below this threshold.
The ratio of forces due to the yield stress and the surface tension leads to a dimensionless number $\tau_c \ell_c / \gamma$.
For a solution of Carbopol, a polymer that is well-known to behave as a Herschel-Bulkley fluid under certain conditions, the typical yield stress is about few tens of Pascals.
Thus, considering $\gamma \simeq 60$ mN/m and $\tau_c \simeq 10$ Pa, we have $\tau_c \ell_c / \gamma \sim 1$ such that we can expect yield stress forces to be significant in dip-coating.

A constitutive equation for yield stress fluids is the Herschel-Bulkley model described by the set of equations

\begin{equation}\label{eq:HB}
    \begin{cases}
        \dot\gamma = 0 & \text{if $\tau<\tau_c$,} \\
        \tau = \tau_c + k \dot\gamma^n & \text{\text{if $\tau>\tau_c$,}}
    \end{cases}
\end{equation}
where $\tau_c$ is the yield stress, $k$ the consistency index, $n$ the flow index and $\dot\gamma$ the shear rate.
Below the yield stress $\tau_c$, the liquid does not flow and above this critical stress, the fluid flows with a stress varying with the shear rate.
The specific case $n=1$, \textit{i.e.} no shear-thinning effect, is referred to as a Bingham fluid and $k$ has the dimension of a viscosity.
The dimensionless number comparing the yield stress to the stress associated to the liquid flow (Eq.~\eqref{eq:HB}) is the Bingham number defined as
\begin{equation}\label{eq:bingham}
    {\rm Bi} = \frac{\tau_c h_0^n}{kV^n}.
\end{equation}

From a qualitative point of view, a yield stress fluid flows at locations where the stress is above the critical stress $\tau_c$.
Therefore, we expect that far from the moving plate, the material remains at rest and in the bulk of the solution, near the plate where the stress is maximum, the material flows across a characteristic thickness that can be coated on the plate.
When the yield stress dominates over the capillary pressure, the thickness of the coated layer must scale as the flowing thickness in the bulk.

As theoretical developments for shear-thinning and viscoelastic fluids are difficult to establish; less studies focused on yield stress fluids as they often combine in practice a yield stress and a shear-thinning effect.
Therefore, analytical solutions are limited.
In 1964, Derjaguin and Levi derived an analytical solution for a Bingham fluid, for an infinitely small viscosity, \textit{i.e.} $k=0$ in Eq.~\eqref{eq:HB} \cite{Derjaguin1964}.
For this constitutive equation corresponding to a \textit{plastic fluid}, the thickness scales as
\begin{equation}\label{eq:YSF_levi}
    \frac{h_0}{\ell_c} \sim \frac{\tau_c^2}{\rho g  \gamma }.
\end{equation}
This equation does not depend on the pulling velocity as the inviscid fluid assumption makes the constitutive equation independent of the shear rate.
From this solution, Derjaguin and Levi generalized to viscous Bingham fluids assuming that the thickness could be written as the sum of a contribution from a Newtonian fluid and a plastic fluid \cite{Derjaguin1964}.

Spiers \textit{et al.} also studied theoretically Bingham fluids \cite{Spiers1975} and they obtained an approximate theory that scales as ${\rm Ca}^{1/6}$.
This result suggests that the film thickness depends weakly on the pulling velocity.
In 1990, Hurez and Tanguy used a finite element analysis to compute the dip coating of Bingham fluids on fibers \cite{Hurez1990}.
They pointed out that the capillary number is not sufficient to describe the flow in the dynamic meniscus and they noticed a \emph{swelling effect} of the meniscus due to the yield stress.
In their simulations, Filali \textit{et al.} noticed that increasing the yield stress increases the coated thickness  \cite{Filali2013}, confirming the swelling effect observed by Hurez and Tanguy.
From these numerical results, they concluded that the film thickness is mainly driven by the flow along the plate in the liquid bath.
A comparison between Newtonian and Bingham fluids is presented in Fig.~\ref{fig:yield_stress_fluid}(a-b) where we clearly see the localization of the shear flow near the plate for the Bingham fluid and a static fluid far from the plate.

Recently, Maillard \textit{et al.} performed experimental \cite{Maillard2014,Maillard2015} and numerical \cite{Maillard2016} studies on the coating of a plate with yield stress fluids.
To obtain a yield stress fluid, they prepared Carbopol solutions at different concentrations.
In these studies, the solutions present yield stress values between $10$ and $100$ Pa and above the yield stress, the solutions are shear-thinning, such that they are well modeled by Eq.~\eqref{eq:HB}.

As pointed out by these authors, some cautions must be taken \cite{Maillard2016}: (i) the fluid must relax after the immersion of the plate and before the withdrawal; (ii) the size of the container must be large enough to avoid wall effects \cite{Maillard2015}; (iii) wall slippage can occur on smooth solids \cite{Bertola2003,Coussot2014}.
On the specific aspect of wall slippage, Maillard \textit{et al.} covered their solid with a waterproof sandpaper that prevents the shear of a less viscous layer at the solid surface.

Once the solid is withdrawn, no drainage of the material is observed.
The shear stress at the solid interface is the maximum shear stress $\rho g h$.
Therefore, for $h<h_c$, the material does not flow by gravity and this condition is  satisfied in their experiments.

Strikingly, the coated thickness does not exhibit a strong dependence with the pulling velocity for yield stress fluids contrary to Newtonian liquids.
Nevertheless, the authors noticed a thickening at large velocities and they rationalized their observations with the Bingham number (Eq.~\eqref{eq:bingham}) as follow.
For small ${\rm Bi}^{-1}$, \textit{i.e.} small pulling velocities, the coated thickness is about $0.3\,\tau_c / (\rho g)$.
For ${\rm Bi}^{-1}>1$, the coated thickness scales as $h_0 \propto V^{n/(n+1)}$.
From these observations, they deduced a general empirical expression for the coated thickness
\begin{equation}\label{eq:YSF_empirical}
    \frac{h_0 }{h_c} = 2 \alpha \left(1 + \left(\alpha^n {\rm Bi}^{-1}\right) \right),
\end{equation}
where $h_c = \tau_c / (\rho g)$.
Despite the agreement between the data and this model, the authors remarked a dispersion of their measurements that is larger than the typical experimental uncertainty \cite{Maillard2016}.
Therefore, some parameters may remain hidden in Eq.~\eqref{eq:YSF_empirical}.
Nevertheless, the model proposed by Derjaguin and Levi for which the coated thickness is proportional to $\tau_c^2$ (Eq.~\eqref{eq:YSF_levi}) can be compared to the empirical Eq.~\eqref{eq:YSF_empirical}.
By definition of $h_c$, the experimental coated thickness scales as $\tau_c$, which is in disagreement with the theoretical prediction \cite{Maillard2016}.

To get further insights in the coating process, the velocity field obtained experimentally from PIV Fig.~\ref{fig:yield_stress_fluid}(c) shows the location of shear.
The shear flow is localized near the substrate  in a layer of constant thickness along the plate and outside this layer, the material is solid.
This thickness increases slowly with the speed of the plate and Maillard \textit{et al.} noticed that this flow is the reverse of the plate immersion.
Furthermore, the coated thickness can be estimated to be about the sheared layer thickness in the bath minus a counter flow due to gravity in the transition domain.
The importance of the flow in the bath could explain the discrepancy in the experimental conditions with the Levi and Derjaguin model that focuses on the flow in the dynamic meniscus \cite{Maillard2016}.
However, we can expect a better agreement for fluids satisfying $\tau_c \ell_c/\gamma \ll 1$, \textit{i.e.} a small perturbation of the yield stress compared to the capillary stress.
This picture of a fluid layer entrainment seems to capture the essential part of the coating mechanism for the studied fluids even if a more details of the flow in the transition domain would be necessary to model the data.

%%%%%%%%%%%%%%%%%%%%%%%%%%%%%%
%
% SOLID
%
%%%%%%%%%%%%%%%%%%%%%%%%%%%%%%

\section{Effect of the solid properties}\label{sec:solid}

In the two previous Sections, we respectively commented on the effects of the liquid-air interface and the bulk rheological properties of the fluid on the coated thickness.
The third key aspect in coatings concerns the properties of the solid, which can be separated in two categories: the surface and the bulk properties.
The surface of the solid material can influence the deposition either by the physical asperities, which modifies the liquid-solid boundary condition from the hydrodynamic point of view, or by the molecular interactions between the liquid and the solid.
Contrary to the effects of the interfacial or the bulk rheological properties of liquids, the mechanics of solid substrates has been less investigated to our knowledge but this topic is gaining more and more attention.
Thus, we present some results that can stimulate future studies.

\subsection{Interfacial effects of solids}

\subsubsection{Rough and textured surfaces}

\begin{figure}
    \includegraphics[width=\linewidth]{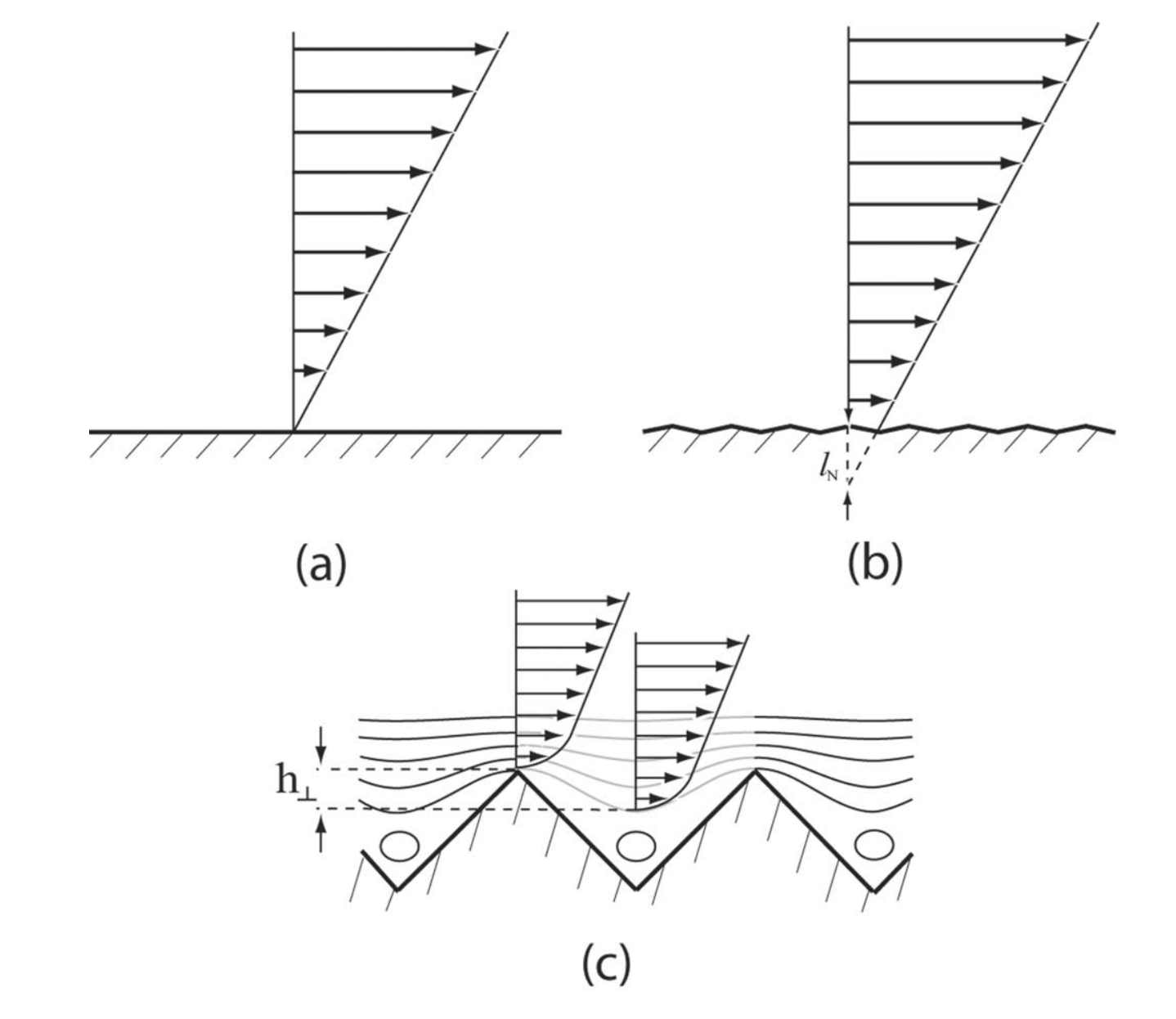}
    \caption{Effect of solid roughness as described by Krechetnikov \textit{et al.} \cite{Krechetnikov2005}.
    Velocity profile on (a) a smooth and (b) on a rough surface where the apparent velocity is larger at the solid interface.
    (c) The vorticity in the grooves generates a slip length.
    The figure is extracted from \cite{Krechetnikov2005}.
    }
    \label{fig:rough_slip}
\end{figure}

Krechetnikov and Homsy studied the effect of surface roughness on the film thickness in a dip-coating experiment \cite{Krechetnikov2005}.
Experimentally, they used sanded glass to produce micron sized grooves characterized afterwards.
For smooth or rough surfaces, below a threshold velocity, the liquid film ruptures and dewets.
First, they observed that the roughness contributes to a stabilization effect of the coated film, \textit{i.e.} the threshold velocity for thickness measurability is lowered.
This effect is attributed to the amplification of the wetting caused by the presence of asperities as described by Wenzel for partially wetting  liquid of contact angle smaller than $90^\circ$ \cite{Wenzel1936}.

The roughness also has a dynamic effect on the film thickness.
For film thicknesses comparable to the average roughness height, experimental measurements indicate a deviation to the Landau-Levich law.
First, the surface roughness produces a thicker film than the one predicted by Landau-Levich (Eq.~\eqref{eq:LLD_plates}).
In addition, the data obtained by Krechetnikov and Homsy scales as
\begin{equation}\label{eq:empiric_roughness}
    h_0 \sim \ell_c \, {\rm Ca}^{0.6},
\end{equation}
where $h_0$ is measured from the top of the grooves.
Krechetnikov and Homsy interpreted the effect of substrate roughness as an enhancement of the production of vorticity, which can be represented by a slip length at the liquid-solid interface (Fig.~\ref{fig:rough_slip}).
From a theoretical analysis, they show that for small slip lengths (\textit{i.e.} small roughness) compared to the film thickness, the film thickness scales as ${\rm Ca}^{2/3}$ whereas for large slip lengths, the film thickness is proportional to the characteristic groove height and independent of the capillary number ${\rm Ca}$.
For the intermediate regime, they do not anticipate a power law with their model and further theoretical and experimental considerations must be address to rationalize this regime.
Therefore, Eq.~\eqref{eq:empiric_roughness} is only established empirically.

\begin{figure}
    \includegraphics[width=\linewidth]{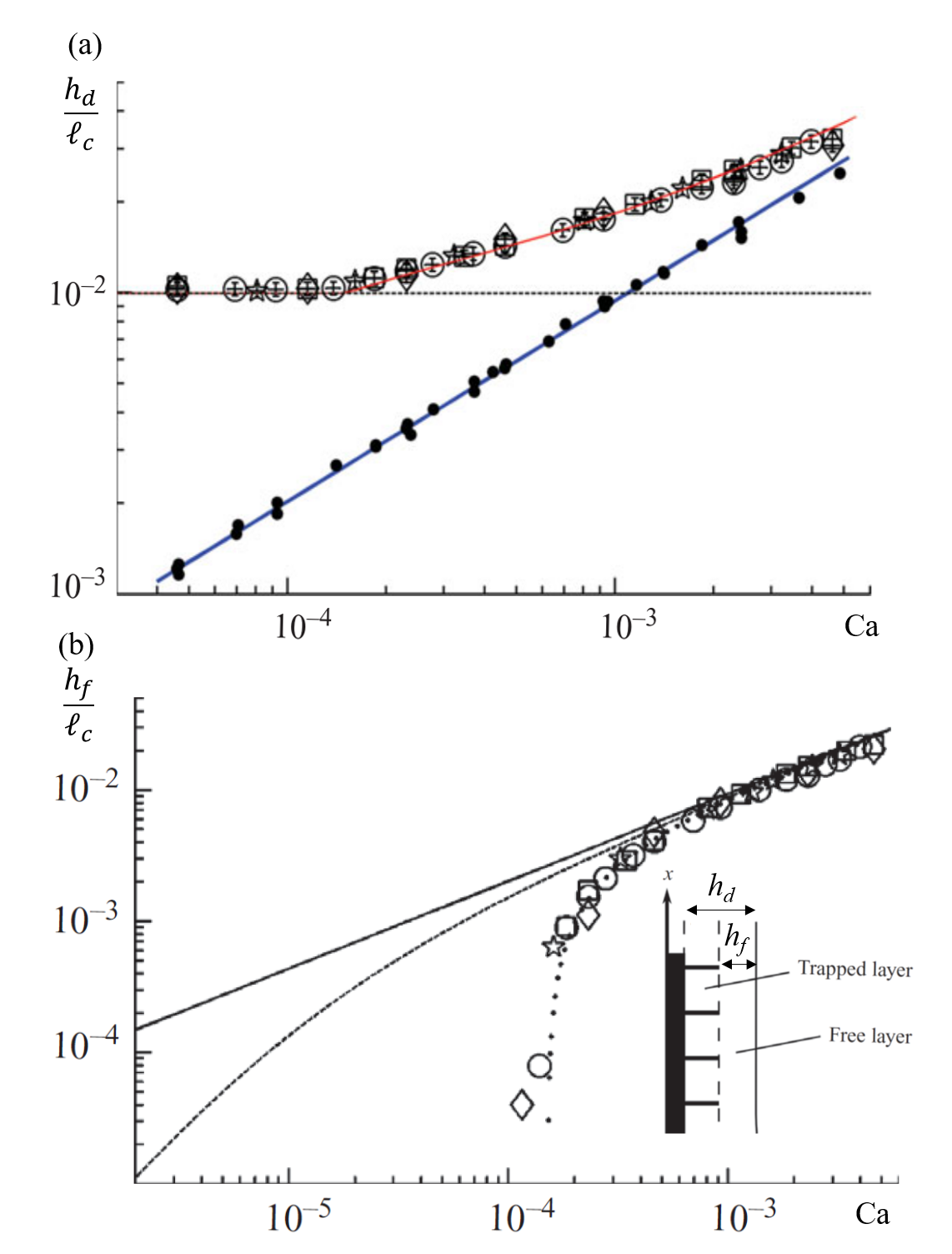}
    \caption{(a) Coated thickness $h_0/\ell_c$ as a function of the capillary number.
    The black dots corresponds to a smooth substrate and the blue solid line represents Eq.~\eqref{eq:LLD_plates}.
    The open symbols are data for textured surfaces where the thickness counts the liquid in the grooves and the free film.
    The horizontal black line is the dimensionless height of the pillars $h_p/\ell_c$ and the red curve is the numerical solution of the adapted Landau-Levich model for a two-layer system.
    (b) Thickness of the free film above the pillars as a function of the capillary number.
    The solid line corresponds to the LDD thickness predicted by Eq.~\eqref{eq:LLD_plates}.
    The inset is a sketch allowing to define $h_f$ and $h_d$.
        The figure is adapted from \cite{Seiwert2011}.
    }
    \label{fig:rough}
\end{figure}

While Krechetnikov and Homsy prepared rough surfaces with a statistical distribution of grooves, Seiwert \textit{et al.} studied a periodic textured surface  \cite{Seiwert2011}.
Rough surfaces are prepared on silicon wafers decorated by a square array of cylindrical pillars of diameter $d=3$ $\mu$m.
The distance between each pillars is 10 or 20 $\mu$m and the pillar height $h_p$ is between 1.4 to 35 $\mu$m.
Coating experiments are carried out with silicon oil and the film thickness is reported in Fig.~\ref{fig:rough}(a) for smooth and textured surfaces.
For smooth surfaces, the Landau-Levich law successfully describes the data and the roughness significantly modifies this behavior.
As for the work of Krechetnikov and Homsy, the entrained film is thicker in presence of roughness.
At low capillary numbers, the film thickness converges to the pillar height $h_p$.
At large capillary numbers, the thickness is the sum of the pillars height $h_p$ and the Landau-Levich thickness $h_{\rm LLD}^p$ as shown in Fig.~\ref{fig:rough}(b).

The difference between the smooth and the textured surfaces is interpreted by Seiwert \textit{et al.} to be caused by a film trapped in the pillars.
Therefore, Seiwert \textit{et al.} proposed a model adapted from the Landau-Levich framework for a two-layer system: a trapped and a free layer.
The resulting differential equation is integrated numerically and leads to a solution, which is not a power law of the capillary number ${\rm Ca}$ contrary to the Landau-Levich theory (Fig.~\ref{fig:rough}).
They predicted the critical capillary number ${\rm Ca}_c$ below which the textured surface entrains a film,
\begin{equation}\label{eq:seiwert}
    {\rm Ca}_c \sim \left( 1+ \frac{h_p^2}{d^2} \right)^{-3/4} \left( \frac{h_p}{\ell_c} \right)^{3/2}.
\end{equation}
In contrast to Krechetnikov and Homsy, Seiwert \textit{et al.} remarked that a slip condition at the surface of the trapped layer does not model the data.

As we mentioned in Sec.~\ref{subsec:YSF}, Maillard \textit{et al.} controlled the interfacial condition of Carbopol solutions, a yield stress fluid, by using a rough substrate to prevent slippage \cite{Maillard2014,Maillard2015}.
This slippage is due to a layer of solvent between the polymers present in the bulk of the solution and the surface of the solid.
Therefore, the roughness of solid surface can either promote or prevent slippage depending on the nature of the liquid.

\subsubsection{Wetting effects}

In the vicinity of small velocities, the film thickness predicted by Landau-Levich-Derjaguin vanishes.
Therefore, in such low capillary number regimes, the Van der Waals interactions responsible for wetting phenomena must be considered \cite{Israelachvili2011}.

First, we describe the case of perfectly wetting liquid.
Van der Waals interactions can be described by a pressure in the thin liquid film.
This pressure is called the \emph{disjoining pressure} \cite{Israelachvili2011}, which can be written as
\begin{equation}
    \pi(h) = - \frac{{\cal A}_{\rm SLV}}{6\pi h^3},
\end{equation}
for a liquid film on a solid where ${\cal A}_{\rm SLV}$ is the Hamacker constant related to the interaction of vapor and solid separated by liquid.
The Hamacker constant ${\cal A}_{\rm SLV}$ is negative resulting in repulsive interactions.

Qu\'er\'e \textit{et al.} reported experimental results on the coating of dodecane on polymeric fibers \cite{Quere1989}.
At small velocities, they observed that the film thickness is constant with a thickness about $0.5$ nm whereas at large velocities, the film thickness satisfies the  Landau-Levich-Derjaguin prediction given by Eq.~\eqref{eq:LLD_fibers}.
Therefore, for fibers, the film thickness can be rationalized as
\begin{subequations}
    \begin{align}
        h_0 &=& \left(  - \frac{{\cal A}_{\rm SLV}\,b}{6\,\pi\,\gamma}\right)^{1/3}, \qquad V < V_0, \label{eq:VdW}\\
        h_0 &=&  h_{\rm LLD}^f  = 1.34 \,b \,{\rm Ca}^{2/3}, \quad V_0 < V \label{eq:LLD_fibers_VdW},
    \end{align}
\end{subequations}
where $V_0 \sim \left( - \frac{{\cal A}_{\rm SLV}}{6\,\pi\,\gamma b^2} \right)^{1/2} \frac{\gamma}{\eta}$ describes the transition below which Van der Waals forces are significant.
Such effect is more difficult to observe experimentally with plates because
the  micrometric size  radius of the fiber is replaced by the capillary length and therefore decreases $V_0$.

\begin{figure}
    \includegraphics[width=\linewidth]{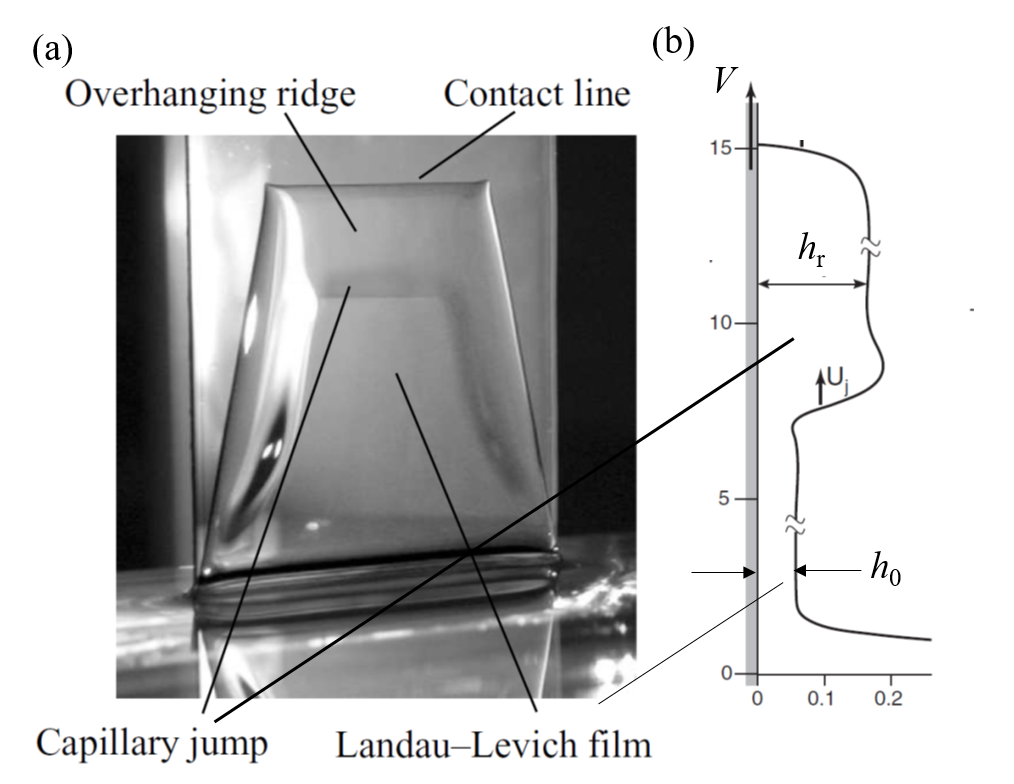}
    \caption{(a) Photograph of a partially wetting liquid (silicon oil) coated on a surface-modified silicon wafer and (b) sketch of the liquid film profile.
    The Landau-Levich film is observed above the dynamic meniscus and a thicker ridge due to dewetting is overhanging.
    The figure is adapted from \cite{Delon2008}.
    }
    \label{fig:dewetting}
\end{figure}

Now, we analyze the coating of partially wetting liquids.
Snoeijer \textit{et al.} revealed that two film thicknesses coexist on the solid as shown in Fig.~\ref{fig:dewetting} \cite{Snoeijer2006,Snoeijer2007,Snoeijer2008,Delon2008}.
For example, in presence of a partially wetting liquid (here silicon oil on a fluorinated substrate), the larger thickness is observed for $ \textrm{Ca} < 10^{-2}$.
To describe the liquid profile, they proposed a model that matches the meniscus of the liquid bath to the Landau-Levich film on one hand, and, on the other hand a solution that matches the contact line at the top of the plate with a second overhanging film.
Then, a capillary jump matches the two flat films and the conservation of mass selects the location of the transition between the thin and the thick films.

Partially wetting liquids changes not only the structure of the coated film.
Below a critical capillary number, a solid withdrawn from a liquid bath is dry in contrast to perfectly wetting  liquids \cite{Quere1998,Quere1999}.
When coated on a solid, a partially wetting liquid undergoes a dewetting phenomena \cite{DeGennes2013}.
Thus, the entrainment of the liquid by a moving solid competes with the receding motion of the contact line.

Derjaguin and Levi assumed that the transition occurs when the dynamic contact angle of the liquid on the solid attains zero \cite{Derjaguin1964}.
However, De Gennes predicted from thermodynamical considerations that the critical contact angle is $\theta_c = \theta_e / \sqrt{3}$, where $\theta_e$ is the equilibrium contact angle \cite{DeGennes1986}.
Therefore, the transition between the dry and the LLD regime is not continuous in terms of contact angles.
Sedev and Petrov conducted experiments on glass fibers with water and glycerol mixtures \cite{Sedev1991}.
They concluded that the threshold corresponds to a zero contact angle in their experiments in accordance with Derjaguin and Levi \cite{Derjaguin1964}.
However, experiments on sliding drops suggest that this angle is non-zero \cite{Rio2005}.

Maleki \textit{et al.} also studied the contact line dynamics of silicon oils on glass cylinders for which the surface is treated chemically to obtain a partial wetting conditions \cite{Maleki2007}.
They found a critical angle $\theta_c$ to be a fraction of $\theta_e$, reminiscent of the prediction of De Gennes \cite{DeGennes1986}.
However, a Cox-Voinov model applied to these data do not satisfactorily render quantitatively this transition, but only qualitatively.
On similar systems, but replacing the glass cylinders by silicon wafer also treated chemically, Delon \textit{et al.} drawn the same conclusion on the
discontinuous transition of contact angles from a dry to a coated regime \cite{Snoeijer2006,Delon2008}.
As we have seen that partial wetting introduces particular conditions to entrain a liquid film, we consider in the next paragraph the case of elastomeric substrates, which often make partial wetting and can also deform under capillary forces.

\subsection{Toward deformable solids}

\begin{figure*}[!ht]
    \centering
    \includegraphics[width=0.7\linewidth]{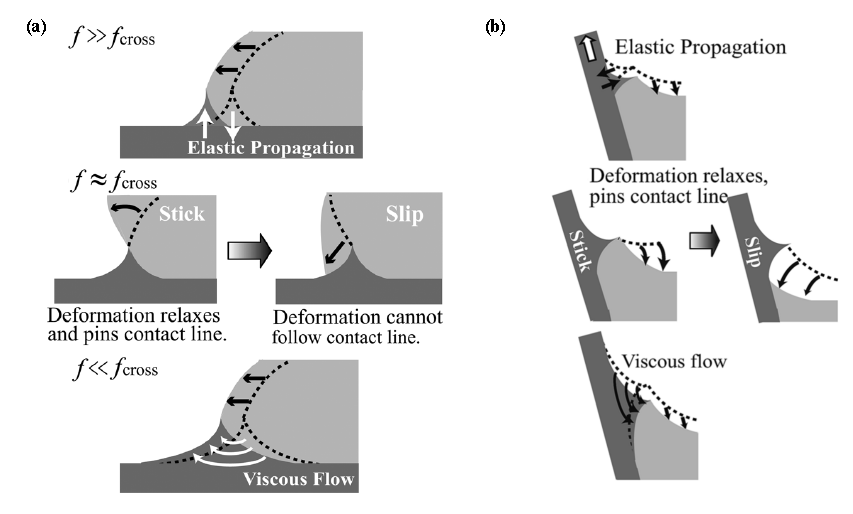}
    \caption{(a) Advancing contact line of a drop on a soft elastomer for three different frequencies compared to $f_{\rm cross}$.
    (b) Configuration of a withdrawn elastomer for the three equivalent frequencies.
    Figure (a) is adapted from \cite{Kajiya2013} and (b) from \cite{Kajiya2014}.
    }
    \label{fig:softsubstrate}
\end{figure*}

Beyond the limit of rigid solids, soft materials have recently gained a large interest since the bulk elasticity of gels and elastomers can interplay capillary forces \cite{Roman2010}.
These systems bring new challenges on the description of soft interfaces and on the coupling between surface and bulk deformations \cite{Andreotti2016}.

When a drop is sitting on a soft material, the gel deforms and makes a ridge to reach the equilibrium with the surface tension (See \cite{Carre1996,Style2012,Lubbers2014,Dervaux2015,Karpitschka2016} for recent works on this subject).
Kajiya \textit{et al.} studied the movement of a contact line on viscoelastic gels by inflating a drop \cite{Kajiya2013}.
The material is mainly viscous at low frequencies smaller than $ f_{\rm cross} = 10^{-3}$ Hz and elastic at higher frequencies.
As shown in Fig.~\ref{fig:softsubstrate}(a), the motion of the contact line of an advancing drop behaves in three distinct regimes: continuous, stick-slip and continuous for increasing characteristic frequencies $f_c = q/(2 \pi R^3)$ where $q$ is the inflation rate and $R$ the drop radius.
At low characteristic frequencies $f\ll f_{\rm cross}$, the deformation of the gel follows the motion of the contact line.
At large frequencies $f\gg f_{\rm cross}$, the gel does not deform significantly as the contact moves rapidly on the surface.
For intermediate frequencies, $f\sim f_{\rm cross}$, the deformation of the gel is concomitant as the contact line advances, which leads to a contact angle hysteresis due to the \emph{asperity} caused by the deformation \cite{Extrand1996}.

Based on this understanding, Kajiya \textit{et al.} analyzed experimentally the motion of the contact line in a dip-coating geometry \cite{Kajiya2014}.
The global dynamic of the contact remains similar to the advancing drop as depicted in Fig.~\ref{fig:softsubstrate}(b).
In addition, they noticed that in the stick-slip regime where the gel is viscoelastic, the stick-slip is regular at the lowest velocities and erratic at larger velocities due to localized pinning spots.
The reverse situation where a thin viscoelastic material bonded on a rigid plate is dipped in a liquid has been investigated by Pu and Severtson \cite{Pu2011}.
Similar regular and irregular stick-slip behaviors has been observed at frequencies corresponding to the viscoelastic regime of the material.

However, to the best of our knowledge, regimes where the liquid is entrained on the soft solid has not been investigated yet.
Considering the tremendous interest for dynamics of wetting on soft solids \cite{Style2013a,Karpitschka2015}, we can expect that this problem will be addressed in a near future.

%%%%%%%%%%%%%%%%%%%%%%%%%%%%%%
%
% CONCLUSION
%
%%%%%%%%%%%%%%%%%%%%%%%%%%%%%%

\section{Perspectives}

In this review, we explored the literature on the coating of plates and fibers by withdrawing the material from a liquid bath.
First, we recalled the theoretical and experimental results for a simple Newtonian liquid on a perfectly wetting and perfectly rigid solid.
Then, we analyzed the existing results through the different parts that constitute the liquid film: the liquid-air interface, the liquid bulk and the liquid-solid interface.
We have shown that each of these domains can bring various complexities of different origins such as the physical-chemistry of surfactants, the rheophysics of complex fluids, the surface properties and the mechanics of the substrate. These complexities have an impact on the boundary conditions, on the flow field and therefore on the final coated thickness.

The first complexity that we explored is the presence of objects at the liquid-air interfaces that influences the coating through the boundary condition at the liquid-air interface.
Different models have been derived corresponding to different limits.
The presence of solid liquid-air interfaces is still almost unexplored despite a pioneering work by Homsy \textit{et al.} \cite{Dixit2013,Ouriemi2013}.
The Gibbs-Marangoni surface elasticity has been taken into account in various models.
The limit of insoluble surfactants successfully describes deviations from the LLD scaling at high capillary numbers \cite{Park1991,Champougny2015}.
Campana \textit{et al} \cite{Campana2010} introduced surfactants solubility, which is promising to explain discrepancies between the LLD prediction and the experiments in terms of prefactors.
Surface viscosity has also been identified as a possible explanation for deviations in terms of prefactor \cite{Scheid2010}.
A quantitative link between the numerical simulations and the chemical parameters is still lacking in most of the situations.

In addition to the interfacial rheology, the rheology of complex fluids makes challenging the comparison between experiments and theoretical predictions.
Indeed, the derivation of hydrodynamic equations becomes sophisticated as constitutive equations bring non-linearities.
The experimental confrontation with these models must be carefully considered since a gap may exist between the constitutive equation and the effective behavior of the fluid, either due to some variations or by additional effects not accounted in the model.
For complex fluids, the flow near solid surfaces can lead to particular boundary conditions due for example to depletion effects for polymer solutions \cite{Barnes1995} or to the particular entanglement structure for polymer melts \cite{DeGennes2003}.

The role of the surface roughness remains unclear, especially regarding the different theoretical approaches of random grooves \cite{Krechetnikov2005} and periodically textured surfaces \cite{Seiwert2011}.
The solid roughness must be considered not only for simple liquids but also for complex fluids where the boundary conditions mentioned previously can be significant on the coated thickness.
Besides the surface properties of the solid, its softness might play a significant effect that remains largely unexplored.

Fluids can be subject to evaporation that not only add a term accounting for the evaporative flux as evaporation can also trigger Marangoni flows due to heat transfer \cite{Qu2002}.
This additional complexity is of importance for coating processes where polymers and colloids are deposited on solid surfaces \cite{Jing2010}.
At low pulling velocities compared to evaporation speed, a stick-slip motion \cite{Abkarian2004} of the contact line reminiscent of the so-called \emph{coffee stain effect} \cite{Deegan1997} is observed while at larger pulling velocities, continuous coatings can be obtained in regimes dominated by LLD \cite{LeBerre2009,Berteloot2013}.
See \cite{Thiele2014} for a recent review.
In the future, it would be fruitful to explore how deposition patterns and coating properties can be tuned by controlling the interfacial or bulk properties that we presented in this review.

\section*{Acknowledgments}

We thank Fr\'ed\'eric Restagno for valuable comments on the manuscript.

        \bibliography{biblio}

        \bibliographystyle{unsrt}

        \end{document}